\documentclass[twocolumn]{aastex631}
\usepackage{amsmath}
\usepackage{amssymb}
\usepackage{threeparttable}
\usepackage{CJK}
\usepackage{booktabs}
\usepackage{tablefootnote}

\def\rg{r_{\rm g}}

\def\kappas{\kappa_{s}}

\def\kappap{\kappa_{P}}
\def\kappar{\kappa_{R}}
\def\taueffp{\tau_{\rm th,P}}
\def\taueffr{\tau_{\rm tr,R}}
\def\rthr{R_{\rm tr,R}}
\def\rthp{R_{\rm th,P}}

\begin{document}
\begin{CJK*}{UTF8}{gbsn}

\title{Multi-band Emission from Star-Disk Collision and Implications for Quasi-Periodic Eruptions}
\author[0000-0003-2868-489X]{Xiaoshan Huang (黄小珊)}
\affiliation{California Institute of Technology, TAPIR, Mail Code 350-17, Pasadena, CA 91125, USA}

\author[0000-0002-8304-1988]{Itai Linial}
\affiliation{Department of Physics and Columbia Astrophysics Laboratory, Columbia University, New York, NY 10027, USA}
\affiliation{Institute for Advanced Study, 1 Einstein Drive, Princeton, NJ 08540, USA}

\author[0000-0002-2624-3399]{Yan-Fei Jiang (姜燕飞)}
\affiliation{Center for Computational Astrophysics, Flatiron Institute, 162 Fifth Avenue, New York, NY 10010, USA}

\begin{abstract}
We perform two-dimensional, multi-group radiation hydrodynamic simulations to explore the observational properties of a solar-like star colliding with an accretion disk around a supermassive black hole at separation of $\sim 100$ gravitational radii. We find that the star-disk collision produces ejecta on both sides of the disk. As the ejecta expand and cool, transient flares arise, reaching peak bolometric luminosity of up to $L\gtrsim10^{43}\rm erg~s^{-1}$. We estimate that the typical light curve rises and decays on an hour timescale. The spectral energy distribution (SED) peaks in $20-50$eV. The optical depth in soft X-rays is lower than the frequency-integrated optical depth, yielding $100$eV-$1$KeV luminosity $\nu L_{\nu}\gtrsim10^{42}\rm erg~s^{-1}$. The ejecta aligned with the star's incident direction shows breakout emission, leading to asymmetric SED evolution of the two ejecta. The SED evolution is roughly consistent with those seen in short-period quasi-periodic eruptions (QPEs), which have eruption duration ranging from sub-hour to hours, but the ejecta cooling emission alone may not be sufficient to explain the longer duration flares. Increasing incident velocity generally produces a brighter and harder flare. A larger disk scale height prolongs the breakout emission but leads to a somewhat softer SED. A higher disk surface density can lead to higher ejecta temperature, reducing bound-free opacity and increasing luminosity. When lowering the disk surface density, we find that the ejecta becomes optically thin when the scattering optical depth across disk is at the order of $\tau_{\rm disk}\sim200$, and the ejecta disappear when $\tau_{\rm disk}\sim10$. 
\end{abstract}


\section{Introduction}\label{sec:introduction}

Most galactic nuclei harbor a supermassive black hole (SMBH) engulfed by a dense nuclear stellar cluster, containing millions to billions of stars and stellar remnants. The interaction between stars and the SMBH or its environment can give rise to a variety of high-energy phenomena, including Tidal Disruption Events (TDEs; \citealt{Rees1988}), gravitational wave-driven inspirals of compact objects or main-sequence stars (Extreme Mass Ratio Inspirals, EMRIs; \citep{Babak_2017,Linial2017,linial2023unstable,Linial2024circularTDE,rom2024dynamics}), ejection of hypervelocity stars \citep{Hills_1988,Brown2015HVS,Verberne2024HVS} and direct collisions between stars or stellar remnants \citep[e.g.,][]{sari2019tidal,Rose2023,Balberg2023}.

Stellar objects in galactic nuclei may also commonly interact with accretion flows near the SMBH. Such encounters are ubiquitous in Active Galactic Nuclei (AGN), where an extended accretion disk is present, overlapping with a fraction of the stellar orbits \citep[e.g.,][]{xian2021x,Wang2024}. Gravitational and hydrodynamical torques from the disk can reshape the orbital distributions, bringing some stars into alignment with the disk, exciting or dampening eccentricities, or possibly ablating the outer layers of disk-traversing stars \citep{yao2025star}. Even in otherwise quiescent galaxies, the occasional disruption of a star as a TDE gives rise to a compact accretion flow evolving on years-decades timescales \citep{Rees1988,Phinney_1989}. If stellar or compact-object EMRIs are present at separations commensurate with the physical size of the newly formed TDE accretion disk, they undergo repeated collisions with the disk on scales of tens to hundreds of gravitational radii \citep{linial2023emri+,franchini2023quasi,tagawa2023flares}.

Repeated star-disk collisions may produce bright recurring flares in galactic nuclei. Quasi-Periodic Eruptions (QPEs) \citep{miniutti2019nine, giustini2020x, arcodia2021x, chakraborty2021possible, quintin2023tormund,
nicholl2024quasi, chakraborty2025discovery,HernandezGarcia2025} are a newly discovered class of soft X-ray flaring sources in relatively low-mass galactic nuclei \citep{Wevers2022}, with typical recurrence timescales ranging from a few hours to a few days. The flare duration is $\sim10-20\%$ of the recurrence time between flares, ranging from tens of minutes to several hours \citep[e.g.,][]{chakraborty2024testing}. The QPE flare luminosities are observed at $10^{41-43}\rm \, erg~s^{-1}$, with SEDs resembling blackbody emission of temperature $k_{\rm B}T\sim 100-200$ eV. QPE sources are often characterized by a quiescent X-ray emission component observed in between eruptions, which is dimmer and cooler than the QPE flares, with luminosities of $10^{40-41} \, {\rm erg \, s^{-1}}$ and blackbody temperatures of $k_{\rm B} T \sim 50 \, {\rm eV}$, broadly consistent with the emission expected from a bare radiatively efficient disk accreting at roughly 1-10\% of the Eddington rate \citep{miniutti2019nine,Miniutti2023_models}. There is now mounting observational evidence that at least some QPEs appear in the aftermath of TDEs, months or years after the detection of a bright flare in optical transient surveys \citep{nicholl2024quasi,chakraborty2025discovery,Quintin2023,Bykov2024}. The emission properties of QPEs, such as their timing patterns and quiescent emission have been shown to gradually evolve over timescales of years \citep{miniutti2023alive,arcodia2024ticking,chakraborty2024testing,pasham2024alive,Miniutti2025}. Various physical scenarios have been proposed to explain the origin of QPEs, including accretion disk instabilities \citep[e.g.,][]{miniutti2019nine,raj2021disk,pan2022disk,kaur2023magnetically}, modulated accretion from Roche lobe overflow of a stellar object \citep[e.g.,][]{zalamea2010white,king2020gsn,king2022quasi,zhao2022quasi,metzger2022interacting,krolik2022quasiperiodic,linial2023unstable,lu2023quasi}, or repeated collisions between a stellar object and an accretion disk \citep{linial2023emri+,franchini2023quasi,tagawa2023flares,xian2021x,yao2025star,vurm2025radiation}. Among these, the latter class of models are potentially better suited to account for an important clue: a regular alternating timing pattern has been observed in some QPE sources: GSN 069, RXJ1301.9+2747, eRO-QPE2 and eRO-QPE4 \citep{miniutti2019nine,giustini2020x,arcodia2021x,arcodia2024more}, where the intervals between consecutive flares alternate back and forth, with variations on the order of $\sim$10\%. This striking behavior could be explained if the flares originate from collisions between a star on an inclined, modestly eccentric orbit and a geometrically thin accretion disk. More complex timing patterns seen in some sources \cite[e.g.,][]{giustini2020x,arcodia2022ero1,HernandezGarcia2025} are often attributed to apsidal and nodal precession of the orbit and the disk \citep[e.g.,][]{franchini2023quasi,chakraborty2024testing,zhou2024probing,xian2025secular}, with alternative scenarios invoked to explain complex longer-term timing evolution \citep[e.g.,][]{arcodia2024ticking,Miniutti2025,linial2025qpes}.

Along with the aforementioned theoretical frameworks, star-disk encounters have been explored in several numerical simulations in the past. Roughly speaking, the relevant numerical aspects include the dynamical interaction and the emission process. A minimum set-up for studying the (hydro-)dynamics involves a compact object or a star that (repeatedly) enters a pre-existing disk. For example, \citet{ivanov1998hydrodynamics} studied the dynamics of a companion black hole ploughing through the accretion disk of an SMBH, finding two-sided plumes of ejecta, above and below the plane of the disk. \citet{lam2025black} studied the outcome of an intermediate mass black hole (IMBH) interacting with an SMBH disk, and found similar ejecta formation and substantial accretion onto the orbiting IMBH. \citet{dodd2025perturbing} studied the global density wave driven by a companion black hole interacting with a disk near the primary black hole, and found consequent episodic accretion rate variations. \citet{ressler2024black} investigated a black hole interacting with pre-existing magnetically-arrested disk and found enhanced outflow as the companion black hole perturbed the disk on an orbit with high inclination.

For SMBH masses of order $10^6 \, M_\odot$ and orbital periods of hours-days, the Keplerian velocities are roughly $(0.05-0.2)\,c$ where $c$ is the speed of light. If the orbiter is a star, its Bondi-Hoyle radius is significantly smaller than its stellar radius, and the interaction is mostly geometrical rather than gravitational \citep{linial2023emri+}. The impact of a supersonic flow traversing a star has been previously studied in the context of a supernova explosion occurring in a binary system \citep[e.g.][]{wong2024shocking,prust2024ejecta}. However, in the QPE context, the ram pressure exerted upon the star as it passes through the disk is significantly weaker than in the case of a nearby supernova explosion, implying that the stellar structure is affected less severely. Recently, \citet{yao2025star} studied star-disk collisions numerically, with an emphasis on stellar ablation. They found that a small fraction of the envelope is removed from the star following repeated disk passages, and that the consequent interaction between the disk and stripped stellar atmosphere can be a promising source to contribute to QPE flares. \cite{linial2025qpes} recently presented an analytical framework for studying the tidal evolution and emission arising from ablated stellar material colliding with the disk, highlighting its possible role in setting QPE emission properties.

For the emission, assuming the photons are produced during the orbiter interacting with disk, regardless the origin of disk and the orbiting companion, the production of soft X-ray photon is essential in order for QPE flares to outshine the quiescent disk. The photon production is likely to emerge from a radiation-mediated shock in the disk, and the hot photons are then be re-processed by the optically thick environment to lower energy band. \citet{linial2023emri+} proposed that due to the limited rate of photon production in the shock downstream (likely dominated by free-free emission), as well as the short disk crossing time, the resulting spectral energy distribution (SED) may deviate from a thermalized black-body radiation. Given this picture, modeling absorption opacities such as free-free (Bremsstrahlung and Compton Scattering), bound-free (photo-ionization) and potentially bound-bound (line opacity), is essential for properly obtaining the flare light curve and time-dependent spectra, and for capturing deviations from blackbody emission. Recently \citet{vurm2025radiation} have studied the spectral evolution from a simple model of the expanding ejecta following a shock driven by a star crossing a disk, and found that soft X-ray emission may arise following such encounters. They have quantified the importance of Compton scattering and demonstrated its potential role in shaping the SED for high-velocity shock ($\gtrsim 0.15 \, c$).

In this work, we explore the dynamics and emission of a star colliding with a section of optically thick accretion disk by two main sets of two-dimensional radiation hydrodynamic (RHD) simulations with Athena++ \citep[][]{stone2020athena++, Jiang2021,jiang2022multigroup}. The first set focuses on the dynamics and spans a larger parameter space. We solve the frequency-averaged radiation transfer equations coupled with the fluid equations, hereafter referred to as the ``gray RHD simulations''. A second set of simulations utilizes the multi-group radiation transfer module to obtain the SED evolution together with dynamics, hereafter we refer them as the ``multi-group simulations''. The gray RHD simulations set the initial condition for the multi-group simulation, with both sets evolving gas dynamics with radiation independently. The effects of three-dimensional dynamics will be explored in a future work. 

The paper is organized as follows: In Section~\ref{sec:analytical} we briefly review the analytical framework and results laid out in \citet{linial2023emri+} in the context of star-disk collisions and QPE flares. The summarized framework is simplified to compare the simulations in this work, readers are referred to \citet{linial2023emri+} to find more complete derivations. In Section~\ref{sec:method_set-up}, we introduce the numerical set-up used in this work. We discuss the gray RHD simulations in Section~\ref{sec:result}, where we show that the generic feature of the bow shock formed in the disk (Section~\ref{result:bow_shock}) and the subsequent ejecta evolution (Section~\ref{subsec:result_ejecta}-Section~\ref{subsec:result_ejectacooling}). We explore the effects of varying the disk surface density, star's incident velocity and disk scale height (Section~\ref{subsec:restult_densdisk}-Section~\ref{subsec:restult_hdisk}). The opacity structure of ejecta and the emission processes are discussed separately in Section~\ref{subsec:result_Opacity}. 

The multi-group simulation results are presented in Section~\ref{sec:result_multi}. Following the previous section, we firstly discuss the multi-group opacity we adopt in the simulation from TOPs opacity data \citep{iglesias1996updated,colgan2016new} in Section~\ref{subsec:multgroup_opacity}. We discuss the SED evolution from these simulations, with a focus on the SED shape resulting from the multi-group opacities in Section~\ref{subsec:multigroup_fiducial}. Then we explore the potential effect of Compton scattering and the quasi-steady state SED with a post-process approach in Section~\ref{subsection:multi_compton}. We vary the disk properties, incident velocities and discuss their effect on SED in Section~\ref{subsec:multi_params}. To explore a different limit and for completeness, we also perform one case study of a star colliding with a low optical depth, puffy disk in Section~\ref{subsec:optically_thin_disk}. According to these results, we discuss favorable conditions for producing short-duration QPEs in Section~\ref{subsec:favored_condition}, including a discussion of caveats and future improvements. We summarize the main findings and conclusions in Section~\ref{sec:summary}.

\section{Brief Summary of Analytical Picture}\label{sec:analytical}
In this section, we briefly summarize the framework of star-disk collisions from \citet{linial2023emri+} and outline the estimates relevant to emission and dynamics in this work. 

When a star passes a disk, it impacts a cylindrical region with cross-section $\sim \pi R_{*}^{2}$ and height $H_{\rm disk}$, where $R_{*}$ is stellar radius and $H_{\rm disk}$ is disk scale height. To the first order, the ejecta mass relates to disk surface density by 

\begin{equation}\label{eq:mej}
    \begin{split}
    M_{\rm ej}&\sim \pi R_{*}^{2}\Sigma_{\rm disk}\\
    &=7.6\times10^{-7}M_{\odot}\bigg(\frac{\Sigma_{\rm disk}}{10^{5}\rm g~cm^{-2}}\bigg)\bigg(\frac{R_{*}}{R_{\odot}}\bigg)^{2}
    \end{split}
\end{equation}

The ejecta velocity is roughly at the order of star incident velocity $v_{\rm ej}\sim v_{*}\sim0.1c$. We estimate the bolometric luminosity from ejecta cooling by assuming that the kinetic energy of shocked disk material $E_{\rm ej}$,  is radiated over the estimated diffusion timescale $t_{\rm diff}$, through a spherical, isotropically expanding ejecta with constant velocity $v_{\rm ej}$, and constant, scattering-dominated opacity. 
The ejecta kinetic energy can be estimated as:
\begin{equation}\label{eq:Eej}
\begin{split}
    E_{\rm ej}&\approx \frac{1}{2}M_{\rm ej}v_{\rm ej}^2\\
    &=6.8\times10^{45}\rm erg\bigg(\frac{\Sigma_{\rm disk}}{10^{5}\rm g~cm^{-2}}\bigg)\bigg(\frac{R_{*}}{R_{\odot}}\bigg)^{2}\bigg(\frac{v_{\rm ej}}{0.1c}\bigg)^{2}
\end{split}
\end{equation}

The photon diffusion timescale can be estimated as follows, assuming the ejecta is a sphere expanding with velocity $v_{\rm ej}$:
\begin{multline}\label{eq:tdiff}
    t_{\rm diff}\approx \bigg(\frac{\kappa M_{\rm ej}}{4\pi cv_{\rm ej}}\bigg)^{1/2}
    \approx 11 \,{\rm min} \; \bigg(\frac{R_{*}}{R_{\odot}}\bigg)\bigg(\frac{\Sigma_{\rm disk}}{10^{5}\rm g~cm^{-2}}\bigg)^{1/2} \\\bigg(\frac{v_{\rm ej}}{0.1c}\bigg)^{-1/2}
    \bigg(\frac{\kappa}{0.32\rm cm^{2}~g^{-1}}\bigg)^{1/2},
\end{multline}
where $\kappa$ is the ejecta opacity. The luminosity is roughly $L_{\rm diff}\sim E_{\rm ej}(V_{\rm init}/4\pi R_{\rm diff}^{3})^{\gamma-1}/t_{\rm diff}$, where we adopt $\gamma=4/3$ for radiation pressure dominated gas. The factor multiplying $E_{\rm ej}$ accounts for adiabatic loss of energy, in which $V_{\rm init}\approx\pi R_{*}^2H_{\rm disk}/7$ estimates the initial volume of shocked disk gas, $4\pi R_{\rm diff}^{3}$ estimates the expanded ejecta size, and $R_{\rm diff}\approx v_{\rm ej}t_{\rm diff}$. This leads to:
\begin{equation}\label{eq:ldiff}
\begin{split}
    L_{\rm diff}&=\frac{E_{\rm ej}}{t_{\rm diff}}\left[\frac{\pi R_{*}^{2}H_{\rm disk}/7}{4\pi(v_{\rm ej}t_{\rm diff})^{3}} \right]^{\gamma-1}=\left(\frac{2}{7}\right)^{1/3}\pi c\kappa^{-1}v_{\rm ej}^{2}R_{*}^{2/3}H_{\rm disk}^{1/3}\\
    &=1.2\times10^{41}\rm erg~s^{-1}\bigg(\frac{v_{\rm ej}}{0.1c}\bigg)^{2}\bigg(\frac{R_{*}}{R_{\odot}}\bigg)^{2/3}\bigg(\frac{H_{\rm disk}}{R_{\odot}}\bigg)^{1/3}
\end{split}    
\end{equation}
where we assume $v_{\rm ej}\sim v_{*}$. We adopt $\kappa=0.32\rm cm^{2}g^{-1}$ and do not explicitly list the dependency on $\kappa^{-1}$ for simplicity.

In the frequency-integrated radiation hydrodynamic simulations, we define thermalization optical depth $\tau_{\rm th,P}$ and photon trapping optical depth $\tau_{\rm tr,R}$ as:
\begin{equation}\label{eq:taueff}
\begin{split}
  &\tau_{\rm th,P}=\int_{r}^{R_{\rm out}}\sqrt{\kappap(\kappap+\kappas)}\rho dr  \\
  &\tau_{\rm tr, R}=\int_{r}^{R_{\rm out}}(\kappar+\kappas)\rho dr
\end{split}
\end{equation}
where $R_{\rm out}$ is the outer edge of the ejecta, $\kappas$ is the constant electron scatter opacity, $\kappar$ is the Rosseland mean opacity and $\kappap$ is the Planck mean opacity. Here we adopt $\kappap$ for absorption opacity in $\tau_{\rm th,P}$, and $\kappar$ for absorption opacity in $\tau_{\rm tr,R}$. We discuss the opacity choice in Section~\ref{sec:method_set-up} and Section~\ref{subsec:result_Opacity}. The corresponding thermalization radius $\rthp$ is the locations where $\taueffp=1.0$, and the photon trapping radius $\rthr$ is set by $\taueffr=c/v$, with $v=v_{\rm ej}$ is ejecta velocity. 

\section{Simulation Set-up}\label{sec:method_set-up}

\begin{table*}
\centering
\caption{Summary of Simulation Parameters}
\label{tab:sim_params}
\begin{threeparttable}
\begin{tabular}{lcccc}
\hline
Name & $\rho_{\rm disk}(\rm g~cm^{-3})$\tnote{1} & $H_{\rm disk}/R_{*}$ & $\tau_{\rm disk,s}$\tnote{2} & $v_{*}/c$ \\ 
\hline
H1\_md\_v0.1c & $2.4\times10^{-7}$ & 1.0 & $5.9\times10^{3}$ & 0.10\\
\hline
H1\_md\_v0.07c & $2.4\times10^{-7}$ & 1.0 & $5.9\times10^{3}$ & 0.07\\
\hline
H1\_md\_v0.15c & $2.4\times10^{-7}$ & 1.0 & $5.9\times10^{3}$ & 0.15\\
\hline
H1\_hd\_v0.1c & $2.4\times10^{-6}$ & 1.0 & $5.9\times10^{4}$ & 0.1\\
\hline
H1\_ld\_v0.1c & $2.4\times10^{-8}$ & 1.0 & $5.9\times10^{2}$ & 0.1\\
\hline
H3\_md\_v0.1c & $7.87\times10^{-8}$ & 3.0 & $5.9\times10^{3}$ & 0.1\\
\hline
H0.3\_md\_v0.1c & $7.87\times10^{-7}$ & 0.3 & $5.9\times10^{3}$ & 0.1\\
\hline
\end{tabular}
\begin{tablenotes}
\item[1] The density at disk center is defined as $\rho_{\rm disk}=\Sigma_{\rm disk}/H_{\rm disk}$ \item[2] The estimated scatter optical depth of disk is defined as $\tau_{\rm disk,s}=\kappa_{\rm es}\Sigma_{\rm disk}$
\end{tablenotes}
\end{threeparttable}
\end{table*}

We perform seven radiation hydrodynamic simulations with frequency-integrated radiation transfer (gray RHD), and six radiation hydrodynamics simulations with multigroup radiation transfer (multigroup), we discuss their set-ups and results in Section~\ref{sec:result} and Section~\ref{sec:result_multi} respectively. The RHD gray simulations and multigroup simulation are using the explicit radiation transfer module in Athena++ \citep{Jiang2021,jiang2022multigroup}. 

We model the process of a star traversing a disk at the local Keplerian velocity as the collision between a gravitating, polytropic star and a column of optically-thick gas. We perform the simulation in the comoving frame of the star, so the disk column moves towards the star with velocity $v_{*}$. Since the involved velocities are on the order of $0.1c$ and well below $0.5c$, we adopt the Newtonian velocity transformation for simplicity. In the rest of the paper, we refer to the frame of simulation as the \textit{comoving frame}, and the frame where the disk has zero bulk velocity as \textit{disk frame}.

The star is assumed to be a $n=3/2$ polytrope for simplicity, the density and pressure profiles are shown in Appendix~\ref{appendix:1Dprofile}. The stellar radius and mass are assumed to be solar. We approximate the self-gravity of star as an explicit gravity source term $g(R)=-Gf(M_{\rm en}(R))$, where $r$ is the distance to the center of star, $M_{\rm en}$ in the enclosed mass at radius $R$, $f(M_{\rm en}(R))$ is a softening potential, and we discuss its implementation in Appendix~\ref{appendix:1Dprofile}. The star's properties and implementation are fixed across all the simulations.

The fiducial disk parameters are inspired by the disk model proposed in \citet{linial2023emri+}, assuming the stellar orbit intersects the disk at $100\rg$ near a $M_{\rm BH}=10^{6}M_{\odot}$ black hole. We model the disk as a column of non-gravitating gas with vertical profile $\rho=\rho_{\rm disk}e^{-z^{2}/2H^{2}}$, where $\rho_{\rm disk}=\Sigma/H$ is the midplane density of the disk given surface density $\Sigma_{\rm disk}$ and scale height $H_{\rm disk}$. The fiducial scale height is comparable to the solar radius $H_{\rm disk}=0.95R_{\odot}$. We choose fiducial surface density to be $\Sigma_{\rm disk}\sim1.7\times10^{4}\rm g~cm^{-2}$, giving $\rho_{\rm mid}=2.4\times10^{-7}\rm g~cm^{-3}$. In the simulations, we vary $\rho_{\rm disk}$ or $H_{\rm disk}$ to change $\Sigma_{\rm disk}$. Gas in the disk has uniform initial temperature $T_{\rm mid}\sim6\times10^{5}$K. Due to high optical depth and short thermal timescale in the disk, radiation and gas will quickly adjust to equilibrium after a few integration cycles, giving disk midplane temperature $T_{\rm mid}\approx2\times10^{5}$K. The effect of disk thermal structure will be explored in future work. The relative velocity between the star and the disk (i.e. the velocity at which the disk is moving towards the star in simulations) varies between $v_{*}=0.07c,~0.1c,~0.15c$ (see Table~\ref{tab:sim_params}). 

In the code, we solve unit-less equations with the scaling of density unit $\rho_{0}=7.87\times10^{-7}\rm g~cm^{-3}$, velocity unit $v_{0}=1.02\times10^{7}\rm cm~s^{-1}$ and length unit $l_{0}=2.21\times10^{11}$cm. In the rest of the paper, we report quantities in c.g.s units unless explicitly specified to be unit-less or other unit. All simulations are performed in two-dimensional Cartesian simulation domain of $[32l_{0}\times50l_{0}]$ in x- and y-direction. The star is located at $x=0,~y=10l_{0}$, the initial disk midplane is at $y_{\rm disk,0}=4l_{0}$. We dye the gas $|y-y_{\rm disk,0}|\leq3.6H_{\rm disk}$ with a disk passive scalar, and dye the gas with $\sqrt{x^{2}+y^{2}}\leq R_{*}$ with a star passive scalar to distinguish gas originated from disk and star. The domain has $[1024\times1600]$ cells in each direction at the root level resolution. We add three-levels of static mesh refinement (SMR) in the center region $[\pm l_{0}\times(10l_{0}\pm l_{0})]$ near the star. With the set-up, we have resolution of $\Delta x=\Delta y=0.1R_{*}$ at the root level, and $\Delta x=\Delta y=0.01R_{*}$ at the highest level. We test using zero to four levels of SMR and found no significant shock structure difference in between level two to four.

For the gray RHD simulations, we adopt the compiled opacity table from \citet{zhu2021global} for Rosseland mean opacity $\kappar$, which extends the OPAL opacity \citep{colgan2016new} to lower temperature limit. We use TOPS opacity for $\kappap$ \citep{iglesias1996updated} and $\kappas=0.34~\rm cm^{2}~g^{-1}$. We discuss the multigroup opacity in Section~\ref{subsec:multgroup_opacity}. The boundary condition for gray RHD simulations are outflow in both directions for hydrodynamic and radiation variables. We adopt density floor $\rho_{\rm floor}/\rho_{0}=10^{-9}$, pressure floor $P_{\rm floor}/\rho_{0}v_{0}^{2}=10^{-11}$, temperature floor radiation transfer module $T_{\rm rad, floor}/T_{0}=2.0\times10^{-4}$. The hydrodynamic variables boundary conditions in all multi-group runs are outflow, and radiation variables are single-direction outflow boundary condition, which copies the outward intensities from the last active zone and sets inward intensities to be zero.

\section{Overall dynamics: Frequency-integrated gray RHD Simulations}\label{sec:result}

\begin{figure*}
    \centering
    \includegraphics[width=0.85\textwidth]{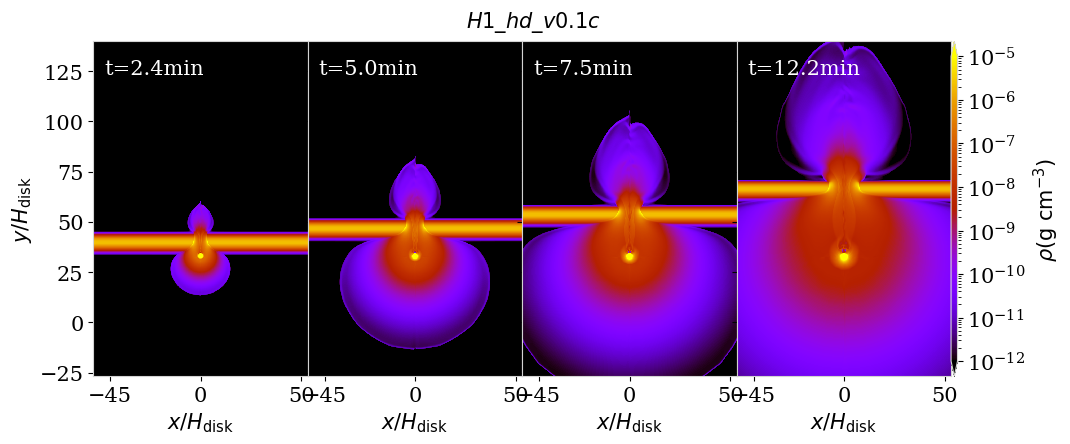}
    \includegraphics[width=0.85\textwidth]{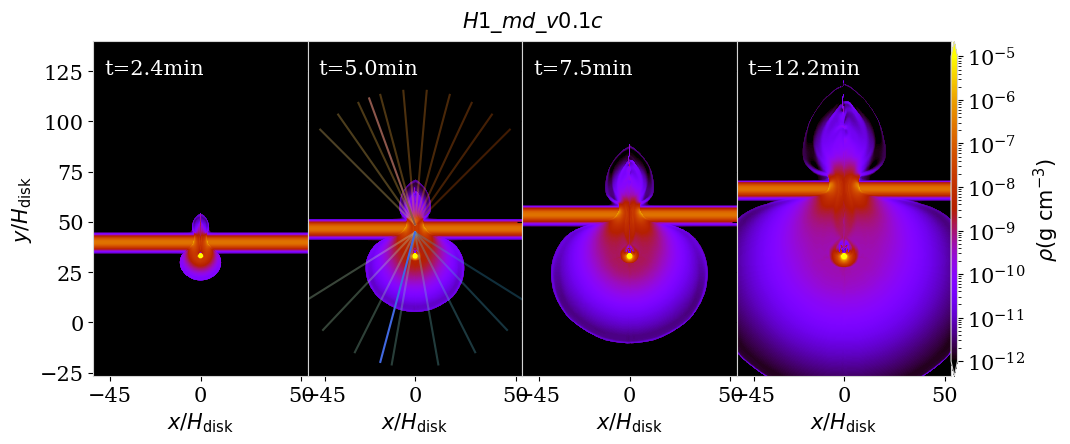}
    \includegraphics[width=0.85\textwidth]{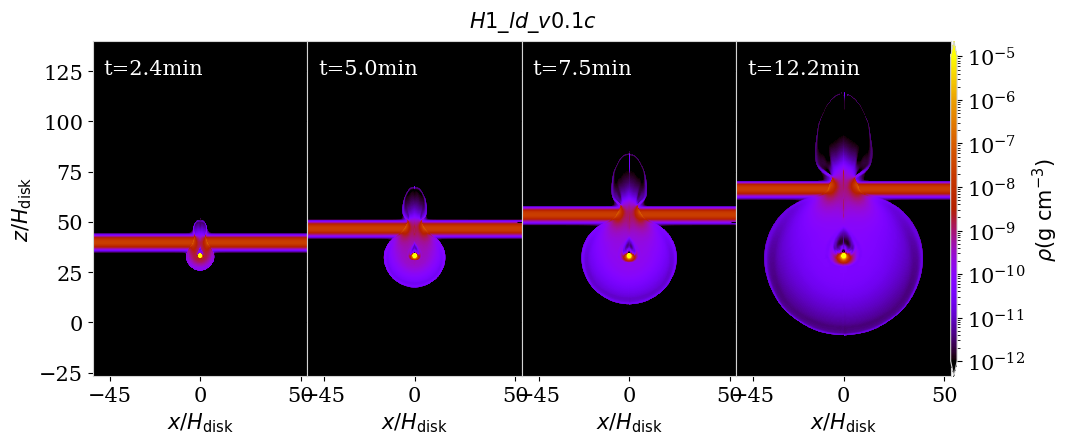}
    \caption{Gas density snapshots from H1\_hd\_v0.1c (the first row), H1\_md\_v0.1c (the second row), H1\_ld\_v0.1c (the third row). The labeled time is the time since the star is at the disk midplane. The colored lines in the second column of H1\_md\_v0.1c corresponds to lines where we sample ejecta 1D profiles. }
    \label{fig:result_density_h1}
\end{figure*}

We present seven simulations with frequency-integrated radiation transfer. We vary disk midplane density $\rho_{\rm disk}$, incident velocity of the star $v_{*}$ and the ratio between stellar size and disk scale height $H_{\rm disk}/R_{*}$, which are listed in Table~\ref{tab:sim_params}.

We refer H1\_md\_v0.1c as the fiducial simulation. For the notation of simulation's name, ``H'' labels the scale height of the disk, H1 corresponds to those $H_{\rm disk}\approx R_{*}$. We vary $\rho_{\rm disk}$ and label them by ``hd'', ``md'', ``ld'', corresponding to high, intermediate and low density. In the fiducial run, the disk and star properties correspond to a quiescent radiative efficient disk around a $M_{\rm BH}=10^{6}M_{\odot}$ black hole that proposed in \citet{linial2023emri+}. The disk is assumed to be accreting at $0.1\dot{M}_{\rm Edd}$, where ${M}_{\rm Edd}$ is the Eddington accretion rate assuming $10\%$ efficiency. We increase (decrease) the $\rho_{\rm disk}$ by one order of magnitude in ``hd'' (``ld'') runs. The ``\_v'' in the simulation name presents $v_{*}$. The star-disk interaction is assumed to happen at $100\rg$, where $\rg=GM_{\rm BH}/c^{2}$, yielding local Keplerian velocity $v_{\rm kep}\approx0.1c$. Because the incident location is not directly set in the simulation, we adopt $v_{*}=0.1c$ as the fiducial incident velocity, and vary the incident velocity between $0.07c,~0.1c,~0.15c$ as parametric study.

\subsection{Bow Shock in the Disk} \label{result:bow_shock}
In H1\_md\_v0.1c, the velocity $v_{*}=0.1c$ corresponds to the Keplerian velocity at the assumed incident location of $100\rg$. The disk is optically thick, with surface density $\Sigma_{\rm disk}\sim10^{4}\rm g~cm^{-2}$ and scale height $H_{\rm disk}=R_{*}$. The second row of Figure~\ref{fig:result_density_h1} shows its gas density snapshots. After the star exits the disk, two plumes of ejecta emerge from both sides of the disk. Hereafter we refer to the ejecta below the disk in Figure~\ref{fig:result_density_h1} as the "forward ejecta", which expands along with the star incident velocity direction, whereas the ejecta above the disk and expands opposites to incident velocity is referred to as the "backward ejecta". 

The star reaches the outer edge of the disk at $y\sim3.6H_{\rm disk}$, it travels through the disk with the increasing density until reaches the disk midplane. The relative velocity between star and disk is supersonic, gas is compressed at the front of the star. A bow shock develops trailing the star, where kinetic energy is dissipated at the shock front and converted into internal energy. Due to the large optical depth, internal energy quickly equilibrate with radiation energy, increasing gas temperature and local radiation energy density. The post-shock gas trailing the star acquires an upward velocity in the disk frame, directed opposite to the star's traveling direction. This oblique bow shock front is essential for producing the backward ejecta.

\begin{figure*}
    \centering
    \includegraphics[width=\textwidth]{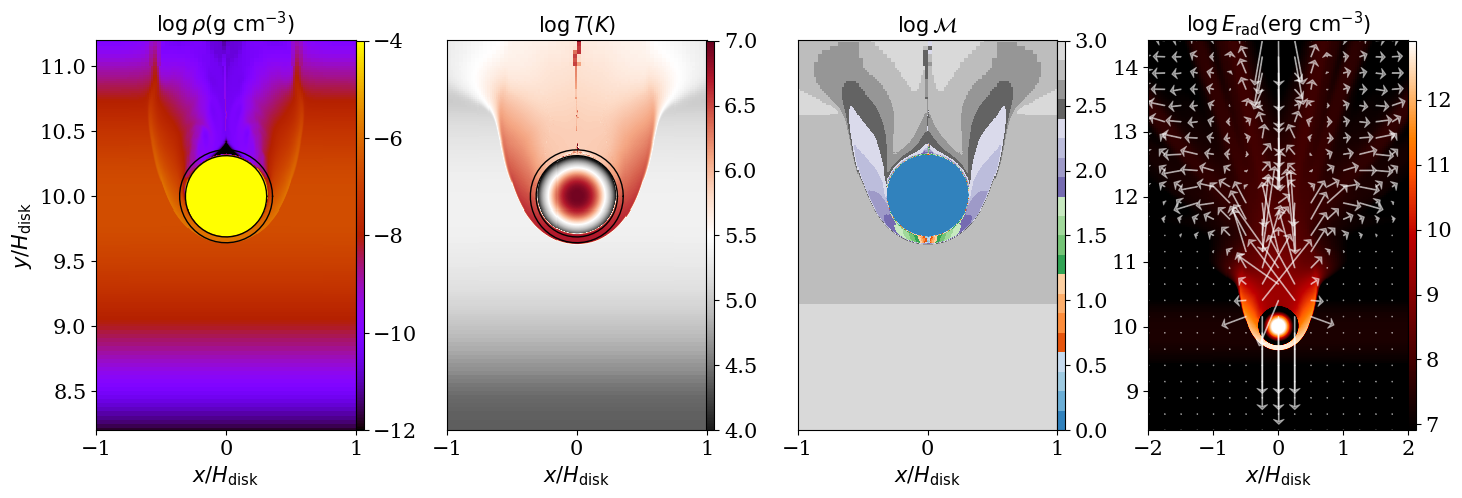}
    \caption{The bow shock formed between the star and disk in H1\_hd\_v0.07c at $t=2.5$ min since the star enters the disk, corresponding to the time when the star is at the midplane of the disk. From left to right, we show gas density, temperature, Mach number and radiation energy density. The two black circles in the first two panels label the stellar radius $R_{*}=R_{\odot}$ and the empirical stand-off distance $1.143R_{*}$. In the rightmost plot, the white arrows show the velocity vectors in the disk frame. }
    \label{fig:star_stand_off}
\end{figure*}

Although we include the gravity of the star, its Bondi radius $r_{\rm Bondi}=2GM_{*}/v_{*}^{2}\approx4\times10^{-4}R_{*}$ is sufficiently small. The formation of backward velocity is different from the scenario that a disk collides with a compact object with Bondi radius $r_{\rm Bondi}\gtrsim H_{\rm disk}$ \cite[e.g.][]{ivanov1998hydrodynamics,dodd2025perturbing}. In the latter case, before the compact object reaches the disk midplane, gas gains backward velocity from the compact object's gravity. The same process repeats when the compact object exits the disk from the midplane and forms forward velocity. In contrast, for the star-disk collision studied in this work, the impact of the star's gravity on disk gas is negligible compared to the bow shock's impact.

Similar problems are studied in the context of planetesimal moving through a proto-planet disk \citep[][e.g.]{mai2020dynamic, yarza2023hydrodynamics}, or a star in a binary that is impacted by its companion supernova explosion \citep{boehner2016imprints, bauer2019remnants, wong2024shocking}. In these applications, a similar bow shock forms in front of a spherical, nearly rigid obstacle embedded in quasi-uniform, constant Mach number fluid. For example, recently \citet{prust2024flow} studied the bow shock formed at the interface of a rigid body in supersonic flow. With numerical simulations, they reproduced shock structures that are consistent with previous laboratory results, including a characteristic length scale: ``stand-off'' distance $R_{\rm stand}$, which describes the distance from the shock nose to the body's boundary. The laboratory experiments in \citet{billig1967shock} founds the fit $R_{\rm stand}/R=1.143$ in the limit of flow Mach number $\mathcal{M}\gg3.24$, where $R$ is the size of rigid body, and the air flow is assumed has $\gamma_{\rm air}\approx7/5$.

In the star-disk collision, $\mathcal{M}=R_{\rm coll}/H_{\rm disk}\sim100 r_{\rm g}/R_{*}$ is at the order of a few hundreds, where $R_{\rm coll}\sim100r_{\rm g}$ is the collision location. The gas is radiation pressure dominated and roughly with effective $\gamma=4/3\lesssim \gamma_{\rm air}$. We assume that in the first interaction of the star, the star is not significantly ablated and does not develop unbound atmosphere \citep[e.g. see a detailed study of stellar atmosphere evolution in][]{yao2025star}, so that the star is approximately rigid. In Figure~\ref{fig:star_stand_off}, we show the empirical stand-off distance relation $R_{\rm stand}\approx1.143R_{*}$ as the larger black circle near the star, the smaller circle corresponds to the initial stellar radius $R_{*}$. 

When the star is at the disk mid-plane, $R_{\rm stand}\approx1.143R_{*}$ approximates the stand-off distance well despite the non-uniform background flow (the disk) and radiation-mediate shock. The gas at the shock front and in the outermost region of the star is heated to $T\gtrsim10^{6}$K, forming a thin region with low Mach $\mathcal{M}\lesssim10$ at the front of the star. The distribution of $\mathcal{M}$ is also qualitatively consistent with the numerical experiment in \citet{prust2024flow}. Far from the shock nose, \citet{yalinewich2016asymptotic} showed that in a uniform background flow, the shape of shock front asymptotes to parabolic as the obstacle plows a quasi-cylindrical cavity in the flow. We find the bow shock front slightly deviates from a parabolic shape, potentially due to the non-uniform background flow. In addition, the disk scale height is only comparable to the star's size, which may further constrain the development of a parabolic wake within the disk.

We show the disk frame velocity vectors as white arrows in the fourth panel of Figure~\ref{fig:star_stand_off}, with the radiation energy density in the disk. The radiation energy is the highest in the central region of the star, where the temperature peaks. Near the surface, the cooler gas forms the peripheral region with lower radiation energy. At the shock front where gas is compressed, radiation energy is significantly enhanced compared to the rest of disk. Shock heated material in the bow shock downstream is swept along by the pressure gradient that develops relative to the lower ambient pressure beyond disk scale height, leaving backward velocity behind the star and forming backward ejecta. 

After the star passing the disk midplane, the bow shock encounters a decreasing density profile, leading to its acceleration in the downward vertical direction \citep[][]{sakurai1960problem}. High radiation pressure pushes disk gas in the direction of shock propagation, leading to the formation of forward ejecta. The acceleration of shock front velocity lead to higher expansion velocity of forward ejecta compared to the backward ejecta. 

Similar to the supernova shock breakout, the bow shock can potentially produce a transient breakout emission from the disk. \citet{tagawa2023flares} found that if the star collides with the disk at a  significant inclination angle, similar breakout emission could be extended as the bow shock sweeps through a larger column of disk material and lead to soft X-ray flares. We discuss this transient breakout emission in Section\ref{subsec:multigroup_fiducial}. The breakout emission and star-disk collision dynamics in the inclined incident will be explored in future simulation works.

\subsection{Evolution of Asymmetric Ejecta}\label{subsec:result_ejecta}

To compare the ejecta morphology with one-dimension, isotopically expanding spherical ejecta, we sample one-dimensional profiles of  hydrodynamical variables along multiple lines through the ejecta. In the second row, second column of Figure~\ref{fig:result_density_h1}, we show a subset of sampling lines. They originate from the disk midplane and extend through the ejecta, along which we measure projected profiles of gas density, velocity and temperature. As the disk moves in the simulation domain, we adjust the starting and ending point of the sampling lines to ensure they originate from same location relative to the disk midplane and maintain the same angle, so that they approximately trace similar part of the ejecta (see Appendix~\ref{appendix:1Dprofile}).

\begin{figure}
    \centering
    \includegraphics[width=\linewidth]{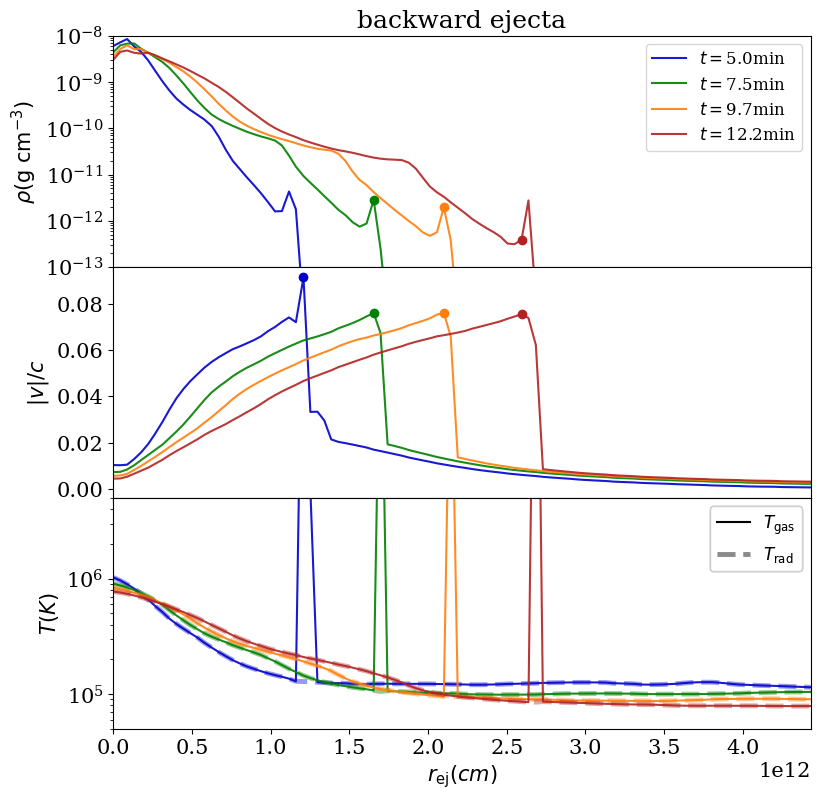}
    \includegraphics[width=\linewidth]{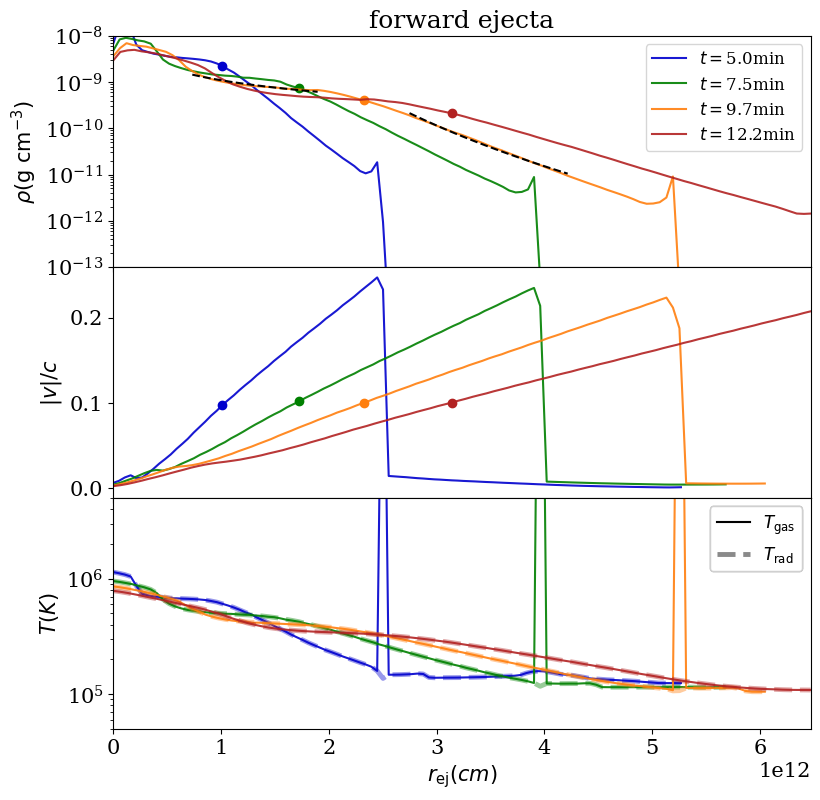}
    \caption{Evolution of ejecta density (the first row), velocity (the second row) and temperature (the third row) profiles of along a representative sampling line. The first plot is for the backward ejecta, the sampling line originates from $2R_{*}$ above the disk midplane with angle $20^{\circ}$ from $x=0$ (the thick brown line in Figure~\ref{fig:result_density_h1}). The second plot is for the forward ejecta, the sampling line originates from $2R_{*}$ below the disk midplane with angle $15^{\circ}$ from $x=0$ (the thick blue line in Figure~\ref{fig:result_density_h1}). The blue, green, yellow, and red lines in the plots correspond to times of 5.0min, 7.5min, 9.7min and 12.2min, respectively, measured since the star crosses disk midplane. The black dashed lines in the first row of  forward ejecta show two power law fits $\rho\propto r^{-0.9}$ and $\rho\propto r^{-7.1}$. The dots in the first and second rows shows the location where projected velocity equals to the star incident velocity $|v_{\rm ej}|=0.1c$. In the upper panel, the blue dot corresponds to $\rho=1.0\times10^{-14}\rm g~cm^{-3}$ and does not appear in the plot.}
    \label{fig:h1_md_mv_lines}
\end{figure}

Figure~\ref{fig:h1_md_mv_lines} shows the evolution of sampled ejecta density, projected velocity and temperature. The density of backward and forward ejecta are comparable, the small density peak at the edge of ejecta is a thin layer of low density gas that is accelerated to high velocity. In H1\_md\_0.1c, the ejecta consist of a dense inner region with a shallower gradient and a low density outer region with a steeper gradient. We fit the forward ejecta density profile by two power laws in Figure~\ref{fig:h1_md_mv_lines}, with a shallower inner region $\rho\propto r^{-0.9}$ and steeper outer region $\rho\propto r^{-7.1}$. Fitting other sampling lines yields similar broken power-law and index range that inner region $\rho\propto r^{-0.9~-~-1.6}$, outer region $\rho\propto r^{-6.5~-~-7.5}$. Similar to supernova ejecta, which are often approximated by a broken power \citep[][]{matzner1999expulsion}, the steeper outer region is from the low density disk gas located farther from the midplane, while the shallower inner region formed from denser disk gas closer to the midplane. Broken power law fits less well to the backward ejecta density profile. 

The exact power law index of density profiles can be sensitive to the dimensionality of the problem, the three-dimensional simulation may give different index than two-dimensional simulation. The morphology will also be affected by the density structure of disk. We assume smooth Gaussian disk density that peaks at midplane, more realistic disk with turbulence and magnetic field may yield more clumpy ejecta.

The forward ejecta velocity can be well-fitted by homologous expansion profile $v\propto r$. The inner high density region roughly corresponds to region where the velocity is below the star's incident velocity $v_{*}=0.1c$ (labeled by the dots in Figure~\ref{fig:h1_md_mv_lines}), while the low density outer region is accelerated to velocities up to $1.5-2$ times the incident velocity. 

The backward ejecta's velocity is significantly slower than the forward ejecta. While the velocity increases from inner region to outer edge of the ejecta, it does not exceed the star's incident velocity. The slower expansion leads to the smaller size of backward ejecta. This asymmetric velocities of backward and forward ejecta originate from the bow shock formed in the disk (Section~\ref{result:bow_shock}). The acceleration of the bow shock when the star travels down the disk density gradient naturally produces higher velocity in the forward ejecta.

The bow shock enhances radiation energy in the disk midplane, the  temperature at the base of backward and forward ejecta are comparable, on the order of $T\sim10^{6}$K. Ejecta temperature decreasing to $\sim10^{5}$K at the outer edge. When the ejecta edge expands to low density ambient gas at local Mach number $\mathcal{M}\sim10^{3}$, a shock forms between the ejecta edge and ambient gas with density $\rho_{\rm amb}=7.9\times10^{-15}\rm g~cm^{-3}$. The shock front is accompanied by a high gas temperature spike known as Zel'dovich spike \citep{1967pswh.book.....Z}, which is not well resolved in our simulations. The radiation temperature $T_{\rm rad}=(E_{\rm rad}/a)^{1/4}$ equals to the gas temperature except for the Zel'dovich spike, suggesting radiation and gas are in thermal equilibrium in the ejecta. Consistently, gas temperature roughly follows $T\propto\rho^{1/3}$, corresponding to optically-thick, radiation pressure dominated gas with effective $\gamma=4/3$.

\subsection{Emission from Ejecta Cooling}\label{subsec:result_ejectacooling}

\begin{figure}
    \centering
    \includegraphics[width=\linewidth]{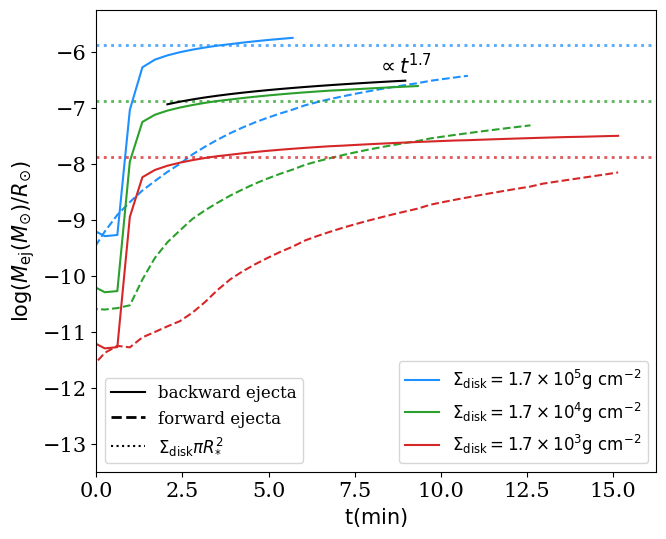}
    \caption{Estimated ejecta mass $M_{\rm ej}$ from H1\_hd\_v0.1c ($\Sigma_{\rm disk}=1.7\times10^{5}\rm g~cm^{-2}$, blue), H1\_md\_v0.1c ($\Sigma_{\rm disk}=1.7\times10^{4}\rm g~cm^{-2}$, green) and H1\_ld\_v0.1c ($\Sigma_{\rm disk}=1.7\times10^{3}\rm g~cm^{-2}$, red). The time axis corresponds to the time since the star is at the disk midplane. For comparison, the Keplerian orbital period at the assumed collision radius is $8.6$ hours. The solid lines shows the $M_{\rm ej}$ below the disk, corresponding to the forward ejecta. The dashed lines show $M_{\rm ej}$ above the disk, corresponding to the backward ejecta. The dotted horizontal lines with corresponding color corresponding to estimation in Equation~\ref{eq:mej} for each simulation. The black solid line shows the power law $M_{\rm ej}\propto t^{1.7}$. The mass $M_{\rm ej}$ is estimated from two-dimensional simulation and  is normalized by $R_{\odot}$. }
    \label{fig:mej_ddisk}
\end{figure}

We estimate the forward (backward) ejecta mass $M_{\rm ej}$ by measuring total mass including all the regions that below (above) $4H_{\rm disk}$ from the disk midplane, and with a disk passive scalar concentration $\geq 90\%$. We also test changing the distance selection criteria to $3H_{\rm disk}$ from disk midplane and did not find orders of magnitude differences. The passive scalar concentration selection criteria is tested from $80\%-95\%$ and also do not affect the result more than order of unity. The measured $M_{\rm ej}$ in H1\_md\_v0.1c is plotted in Figure~\ref{fig:mej_ddisk} as the green lines. Due to the two-dimensional nature of the simulation, we normalize the mass $M_{\rm ej}$ measured from surface density to $R_{\odot}$, the ejecta mass can also be sensitive to the dimensionality of the problem.

The forward ejecta mass increases promptly as the ejecta expands outside the disk in minutes, and then slowly increases for tens of minutes. This later slower increment can be approximated by $M_{\rm ej}\propto t^{1.7}$, shown in Figure~\ref{fig:mej_ddisk}. The backward ejecta mass is smaller than the forward ejecta and increases slower. For long-term ejecta evolution that beyond the simulation domain, we expect the mass increment will slow down when the disk relaxes to its unperturbed state after the collision. The dotted horizontal line corresponds to $M_{\rm ej}$ in Equation~\ref{eq:mej}, which approximates measured $M_{\rm ej}$ within an order of magnitude.

Due to the non-spherical geometry, the ejecta mass derived from two-dimensional results is not consistent with integrating the sampled one-dimensional density profile in Figure~\ref{fig:h1_md_mv_lines} assuming spherical geometry. The latter $M_{\rm ej,sph}=\int_{R_{\rm in}}^{R_{\rm out}}4\pi r^{2}\rho dr$ is higher than the measured ejecta mass in Figure~\ref{fig:mej_ddisk}. The sampled one dimensional profiles of temperature, projected velocity and opacity are relatively insensitive to the non-spherical geometry and can represent the typical values of these quantities in the ejecta. The backward ejecta is more non-spherical. A ``face-on'' line of sight (perpendicular to the disk) results in a higher density column compared to an ``edge-on'' line of sight (parallel to the disk). From the one-dimensional analysis of the counterpart multigroup simulation (which is discussed later in Section~\ref{subsec:multigroup_fiducial}), we find the soft X-ray band flux can decrease up to one order of magnitude when changing line of sight from ``edge-on'' to ``face-on''. Exploring the emission's dependence on geometry and potential viewing angle effect will be discussed in future work.

\begin{figure}
    \centering
    \includegraphics[width=\linewidth]{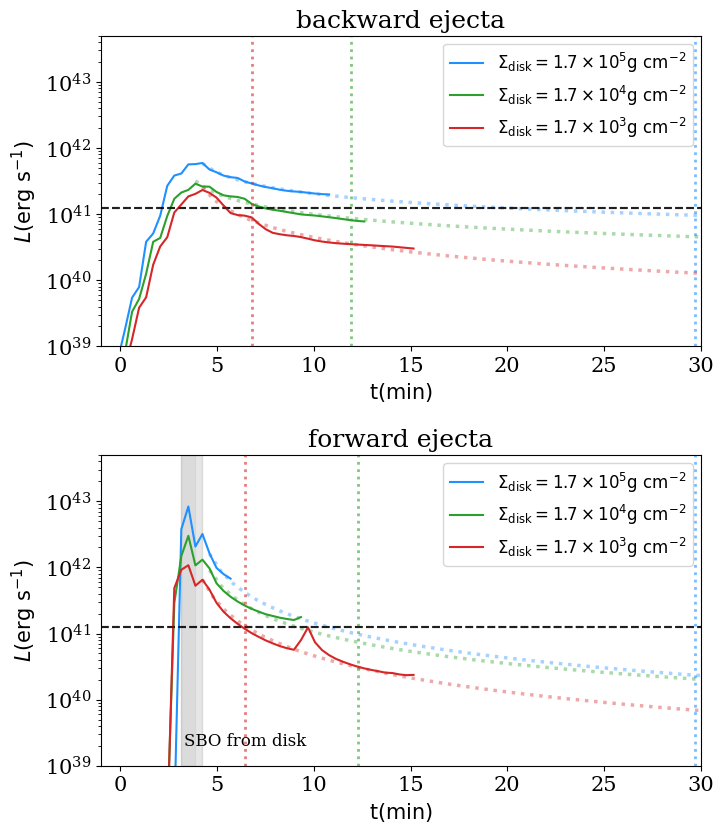}
    \caption{Estimated total luminosity for the backward ejecta (upper panel) and the forward ejecta (lower panel). The blue, green, red lines correspond to H1\_hd\_0.1c ($\Sigma_{\rm disk}=1.7\times10^{5}\rm g~cm^{-2}$), H1\_md\_0.1c ($\Sigma_{\rm disk}=1.7\times10^{4}\rm g~cm^{-2}$), H1\_ld\_0.1c ($\Sigma_{\rm disk}=1.7\times10^{3}\rm g~cm^{-2}$). The time axis corresponds to the time since the star is at the disk midplane. For comparison, the Keplerian orbital period at the assumed collision radius is $8.6$ hours. In each plot, the vertical dotted lines show the estimated diffusion timescale $t_{\rm diff}$ (Equation~\ref{eq:tdiff}). The dotted lines show fitting of power law decay $L\propto t^{-\beta}$. In the lower panel, the gray shaded region marks the brief breakout emission. In the lower panel, the peak of the red line near $t\approx10$ min is due to the numerical effect of angular discretized radiation field.}
    \label{fig:lum_ddisk}
\end{figure}

We estimate the luminosity by measuring the total radiation flux integrated over the boundary surfaces $L=\int \mathbf{F}_{\rm r}\cdot\mathbf{dA}$. For the backward (forward) ejecta, we include only the boundary above (below) the disk midplane. The disk region that extends to the boundary is optically thick and does not significantly contribute to the radiation flux. The measured luminosity is shown in Figure~\ref{fig:lum_ddisk} as the green lines. 

We calculate $L_{\rm diff}$ as in Equation~\ref{eq:ldiff}. For backward and forward ejecta, $L_{\rm diff}=1.23\times10^{41}\rm erg~s^{-1}$, corresponding to the black dashed line in the Figure~\ref{fig:lum_ddisk}. It can be slightly lower than the measured peak luminosity, which is $L_{\rm peak}=2.9\times10^{41}\rm erg~s^{-1}$  for the backward ejecta and $L_{\rm peak}=3.0\times10^{42}\rm erg~s^{-1}$ for the forward ejecta. Interestingly,  the measured luminosity is similar to $L_{\rm diff}$ at one diffusion timescale $t_{\rm diff}$ (the vertical dotted line).

In the forward ejecta, the measured $L_{\rm peak}$ corresponds the transient breakout emission from disk instead of the ejecta cooling emission. We define the duration of breakout emission as the duration between luminosity peaks and when it drops to half of the peak value. It is labeled as the gray shaded region in Figure~\ref{fig:lum_ddisk}. The breakout only lasts less than two minutes, the ejecta cooling emission is the primary source of luminosity in the decay phase. We discuss the spectral energy of breakout emission in Section~\ref{sec:result_multi}.

We show the fittings of $L\propto L_{\rm peak}((t-t_{\rm peak})/\sigma+1)^{-\beta}$ (see also Equation~\ref{eq:expdecay}) as the dotted curve with $\beta=1.33$. Such power law decay format is also adopted in \citep{vurm2025radiation} as generalized form of luminosity decay. The fitting provides a rough estimation of the emission before the entire ejecta becomes transparent, afterwards, the light curve evolves to fast decay as photons escape from the optically thin material rapidly. 

\subsection{Dependency on Disk Surface Density}\label{subsec:restult_densdisk}
We increase (decrease) disk midplane density in the simulations labeled by \_hd (\_ld), so the surface density of H1\_hd\_v0.1c, H1\_md\_v0.1c, H1\_ld\_v0.1c are $\Sigma_{\rm disk}=1.7\times10^{5}\rm g~cm^{-2},~1.7\times10^{4}\rm g~cm^{-2},~1.7\times10^{3}\rm g~cm^{-2}$. The density snapshots are shown in Figure~\ref{fig:result_density_h1}. The ejecta morphology is similar, following the trend that the ejecta density increases as $\Sigma_{\rm disk}$ increases. 

\begin{figure}
    \centering
    \includegraphics[width=0.9\linewidth]{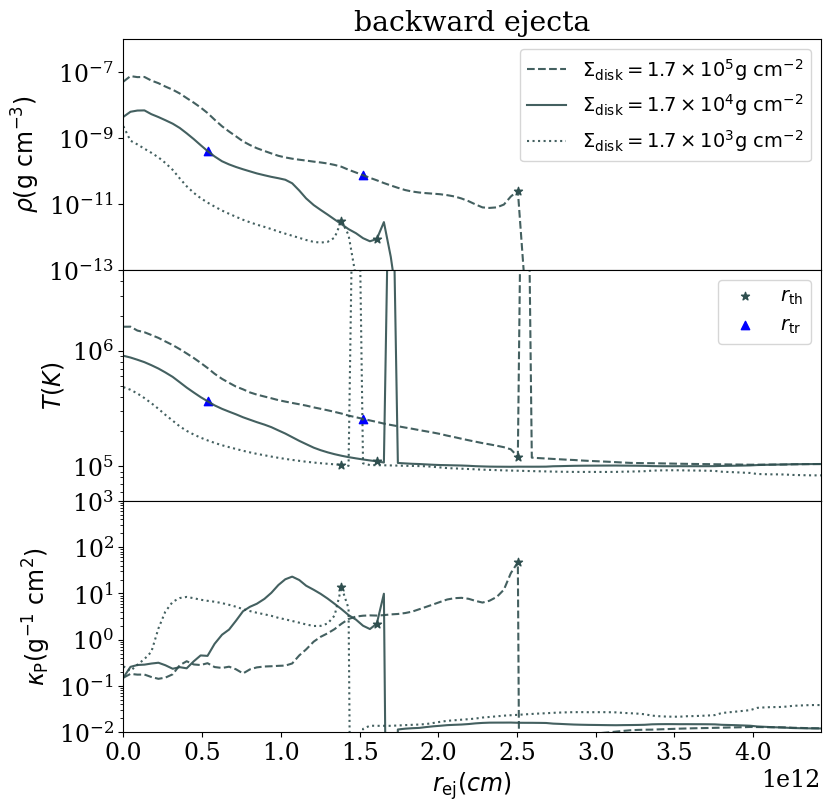}
    \includegraphics[width=0.9\linewidth]{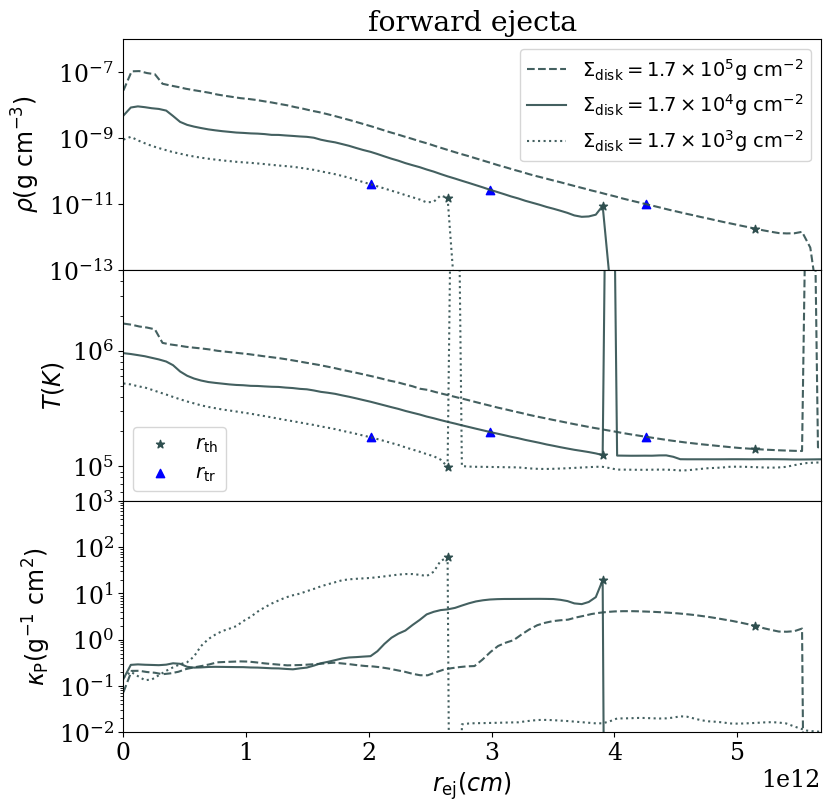}
    \caption{Ejecta density (the first row), temperature (the second row) and Planck mean opacity (the third row) profiles of along the representative lines. In each panel, the dashed, solid and dotted line corresponds to H1\_hd\_0.1c ($\Sigma_{\rm disk}=1.7\times10^{5}\rm g~cm^{-2}$), H1\_md\_0.1c ($\Sigma_{\rm disk}=1.7\times10^{4}\rm g~cm^{-2}$), H1\_ld\_0.1c ($\Sigma_{\rm disk}=1.7\times10^{3}\rm g~cm^{-2}$) at $t=12.4$min. The first plot is for the backward ejecta, the sampling line originates from $2R_{*}$ above the disk midplane with angle $20^{\circ}$ from $x=0$ (the thick brown line in Figure~\ref{fig:result_density_h1}. The second plot is for the forward ejecta, the sampling line originates from $2R_{*}$ below the disk midplane with angle $15^{\circ}$ from $x=0$ (the thick blue line in Figure~\ref{fig:result_density_h1}. The lines are sampled at simulation time 5.0min since the star is at disk midplane. In each panel, the star data points are where $\tau_{\rm eff, P}=c/v$, triangle data points are where $\tau_{\rm eff, R}=c/v$.}
    \label{fig:linecut_ddisk}
\end{figure}

The ejecta mass $M_{\rm ej}$ are shown in Figure~\ref{fig:mej_ddisk} as the blue and red lines. The time-dependent increment is similar to H1\_md\_v0.1c, with $M_{\rm ej}\propto\Sigma_{\rm disk}$ . The forward ejecta increment can be well approximated by $M_{\rm ej,forward}\propto t^{2.2}, ~t^{1.4}$ in H1\_hd\_v0.1c and H1\_ld\_v0.1c. 

In Figure~\ref{fig:linecut_ddisk}, we show one-dimensional profiles as described in Section~\ref{subsec:result_ejecta}. The ejecta density is roughly proportional to $\Sigma_{\rm disk}$. Similar to H1\_md\_0.1c, the forward ejecta density can be approximated by broken power law, with inner and outer region roughly separated by $v_{\rm ej}\approx v_{*}$. The best fit power law is $\rho\propto r^{-0.91}$ and $\rho\propto r^{-5.2}$ in H1\_ld\_v0.1c, $\rho\propto r^{-1.37}$ and $\rho\propto r^{-7.24}$ in H1\_hd\_v0.1c. The velocity can be well-approximated by $v_{\rm ej}\propto r_{\rm ej}$, the inner regions have similar velocity as H1\_md\_v0.1c (Figure~\ref{fig:hdisk03_hydrolines}), the outer region velocity is higher in simulations with larger $\Sigma_{\rm disk}$. But the fast-expanding outer region does not significantly contribute to the total $M_{\rm ej}$. The ejecta temperature shows the trend that higher $\Sigma_{\rm disk}$ yields higher $T_{\rm ej}$, which has interesting implications for ejecta opacity structure.

The third row in each of panel of Figure~\ref{fig:linecut_ddisk} shows the Planck mean opacity $\kappap$. Given the ejecta density range $\rho\lesssim10^{-8}\rm g~cm^{-3}$ and temperature range $T>10^{5}$K, the TOPS Planck mean opacity includes significant contributions from bound-free transitions, yielding $\kappap\gg\kappa_{\rm ff}$. In both backward and forward ejecta, $\kappap\lesssim\kappas$ in the inner region where gas temperature $T\gtrsim5\times10^{5}$K. As temperature decreases towards outer ejecta, $\kappap$ increases above the $\kappas$. It drops again once temperature $T\lesssim2\times10^{5}$K, corresponding to the ``Iron peak'' in opacity range \citep{magee1995atomic,jiang2016iron, piro2024late}. 

The Rosseland mean opacity $\kappar$ can be smaller than $\kappap$ up to $\sim10^{2}$ times, yielding $\kappar\approx\kappas$ in most of the ejecta except for the outermost region, where temperature $T\lesssim2\times10^{5}$K and $\kappar\gtrsim\kappas$. In Figure~\ref{fig:linecut_ddisk}, we label $\rthp$ by the asterisk data points and $\rthr$ (Equation~\ref{eq:taueff}) as the triangle data points. In both backward and forward ejecta, $\rthp$ locate near the ejecta edge, where temperature $T\lesssim2\times10^{5}$K, suggesting efficient energy exchange between radiation and gas in the ejecta.  In H1\_hd\_v0.1c, the ejecta temperature is higher, reducing $\kappap$ and $\taueffp$ and pushing $\rthp$ inside the ejecta edge. Near the Zel'dovich spike, the opacity per simulation cell jumps significantly, which may affect the $\taueffp$ estimations.

The forward ejecta from different $\Sigma_{\rm disk}$ simulations show similar temperature at $\rthr$, where the $T\lesssim1.8\times10^{5}$K and $\kappar+\kappas\approx0.5\rm g^{-1}cm^{2}$, giving $\rthr$ slightly smaller than scatter photosphere. The backward ejecta velocities are generally slower, making $\rthp$ inner than the forward ejecta, so the temperature at $\rthr$ is higher. H1\_ld\_v0.1c has $\taueffr<c/v$ for entire sampling line due to low density, thus we do not label $\rthr$ for it. A common trend is that $\rthr<\rthp$, which is related to $\kappap\gg\kappar\sim\kappas$ and the non-negligible contribution of bound-free opacity in $\kappap$.

Their luminosity is shown in Figure~\ref{fig:lum_ddisk} as the blue and red lines. Increasing $\Sigma_{\rm disk}$ enhances luminosity in both backward and forward directions. The vertical dotted lines are $t_{\rm diff}$ (Equation~\ref{eq:tdiff}), and the horizontal dashed line is the estimated luminosity by Equation~\ref{eq:ldiff}. The spikes in H1\_ld\_v0.1c are due to numerical effect of angular discretization of radiation field and finite rectangular simulation domain. We test increasing the angular grid from 80 angular to 120 angular bins and find a smoother $L$. 

The best fitting parameters for the power law backward ejecta decay indicates slope $\beta=-0.52,~-1.10,~-1.14$, and timescale $\sigma=0.91,~3.32,~1.59$min for H1\_hd\_v0.1c, H1\_md\_v0.1c, H1\_ld\_v0.1c. For the forward ejecta, the best fitting parameters are slope $\beta=-0.43,~-0.67,-1.24$ and $\sigma=0.16,~0.31,~0.80$min for H1\_hd\_v0.1c, H1\_md\_v0.1c, H1\_ld\_v0.1c. The fitting slopes show the trend that higher $\Sigma_{\rm disk}$ luminosity decays slower.

H1\_hd\_v0.1c has the longest $t_{\rm diff}$, but the higher outer region velocity makes the ejecta expand to the simulation domain boundary earlier than others, so $L$ in Figure~\ref{fig:lum_ddisk} is the shortest. The fitting of backward $L$ decays estimates moderate luminosity decay until $t_{\rm diff}$.  The fitted power law of forward ejecta decay suggests they can potentially have luminosity $\gtrsim10^{41}\rm erg~s^{-1}$ for $\sim30$ min. But the fitted evolution may be affected by the relatively short simulation duration. H1\_ld\_v0.1c has the shortest $t_{\rm diff}$ so the measured $L$ covers the first two $t_{\rm diff}$. The backward and forward $L$ drops to $0.38L_{\rm peak}$ and $0.11L_{\rm peak}$ in the first $t_{\rm diff}$, the estimated luminosity by Equation~\ref{eq:ldiff} approximates luminosity at one $t_{\rm diff}$ well. 

In Equation~\ref{eq:ldiff} , the luminosity does not explicitly depend on $\Sigma_{\rm disk}$.  However, the diffusion time estimated in Equation~\ref{eq:tdiff} does depend on $\Sigma_{\rm disk}$. Equation~\ref{eq:ldiff} generally approximates the measured luminosity from simulations at $t_{\rm diff}$ (Figure~\ref{fig:lum_ddisk}, ). We attribute the $\Sigma_{\rm disk}-$dependency of luminosity to the opacity variation in the ejecta. If keep increasing $\Sigma_{\rm disk}$ and the ejecta temperature is high enough that the bound-free opacity is less important and the opacity is dominated by free-free, we speculate the luminosity may be less sensitive to $\Sigma_{\rm disk}$. We discuss ejecta opacity structure in Section~\ref{subsec:result_Opacity}.

\subsection{Dependency on Incident velocity}\label{subsec:restult_starvel}

As the star's incident velocity roughly sets the initial kinetic energy of the shocked disk before ejecta expansion (Equation~\ref{eq:Eej}), we vary the incident velocity between $v_{*}=0.07c,~0.15c$ for the fiducial $\Sigma_{\rm disk}$ in H1\_md\_v0.07c and H1\_md\_v0.15c, as listed in Table~\ref{tab:sim_params}.

The ejecta morphology does not strongly depend on $v_{*}$ and is similar to H1\_md\_v0.1c (Figure\ref{fig:result_density_h1}). The backward ejecta is slower than the forward ejecta, the relative velocity difference in the two ejecta is not sensitive to the tested $v_{*}$. 

\begin{figure}
    \centering
    \includegraphics[width=\linewidth]{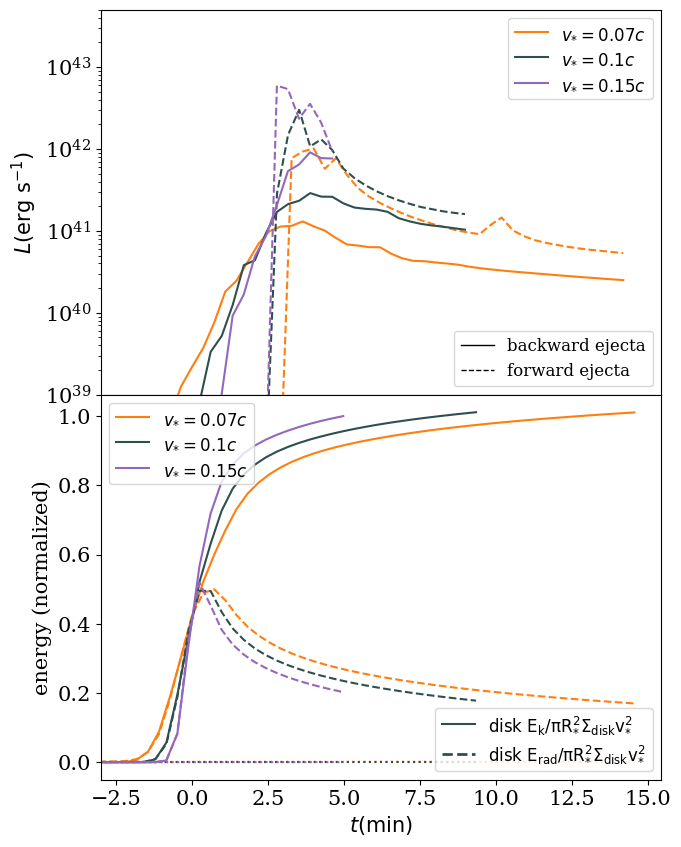}
    \caption{Upper panel: Measured luminosity of backward ejecta (solid lines) and forward ejecta (dashed lines). We stop the luminosity curves when the ejecta expand beyond the simulation domain. The time axis is the time since the star crosses the disk midplane. Lower panel: Normalized gas kinetic energy (solid lines) and radiation energy (dashed lines) for H1\_md\_0.15c (purple, $v_{*}=0.15c$), H1\_md\_0.1c (dark, $v_{*}=0.1c$) and H1\_md\_0.07c (orange, $v_{*}=0.07c$). The normalization factor corresponds to the estimated shocked gas kinetic energy in Equation~\ref{eq:normal_ek_er}. The time delay between the lower panel radiation energy density peak and upper panel luminosity peak is primarily due to photon diffusion time from the shock front to ejecta photosphere.}
    \label{fig:energy_lum_vz}
\end{figure}

The upper panel of Figure~\ref{fig:energy_lum_vz} shows the measured luminosity (similar to Figure~\ref{fig:lum_ddisk}).  The velocity range of $v_{*}=0.07c,~0.1c,~0.15c$ corresponds to a factor of two changes in the star's kinetic energy. We find that the ejecta mass $M_{\rm ej}$ is similar between the three simulations, so the kinetic energy of shocked disk gas is roughly proportional to $\propto v_{*}^{2}$. It is further redistributed to ejecta's kinetic, radiation and internal energy.  The luminosity reflects the fraction of radiation energy diffusing out from ejecta at a given time. The ejecta velocity profiles are homologous regardless of $v_{*}$.

To quantify the energy redistribution, we first transform velocity to the disk frame by $\textbf{v}'=\textbf{v}-\textbf{v}_{*}$, where $\textbf{v}$ is the velocity in the frame of simulation, $\textbf{v}_{*}$ is the incident velocity, i.e. the initial velocity that the disk colliding with the star. The kinetic and radiation energy of disk gas can be normalized as:
\begin{equation}\label{eq:normal_ek_er}
    E_{\rm k}=
    \frac{\int\frac{1}{2}\rho f_{\rm disk} (\textbf{v}-\textbf{v}_{*})^{2} dV}{\pi R_{*}^{2}\Sigma_{\rm disk}v_{*}^{2}},~E_{\rm rad}=\frac{\int e_{\rm rad}f_{\rm disk}dV}{\pi R_{*}^{2}\Sigma_{\rm disk}v_{*}^{2}},
\end{equation}
where $f_{\rm disk}$ is the concentration of passive scalar that marks gas initially in the disk. The volume integration is done in two-dimensional Cartesian simulation domain, with the third dimension thickness normalized to $R_{*}$. We treat the velocity transformation as Newtonian due to relatively low $v_{*}$. 

The normalization is chosen to be estimated kinetic energy of shocked gas $E_{\rm k,ej}$ (Equation~\ref{eq:Eej}). The volume integration includes forward and backward ejecta, so we account for a factor of two when estimating ejecta mass using the star's cross-section $\pi R_{*}^{2}$, considering the relative motion of star and disk \citep{linial2023emri+}. Equation~\ref{eq:Eej} is not identical to the initial shocked gas kinetic energy in simulation, nor does it sets energy up-limit, so the normalized energy can be larger than unity. 

In the lower panel of Figure~\ref{fig:energy_lum_vz}, a small fraction of the star's kinetic energy imparts to the disk during the star-disk collision, the kinetic and radiation energy in the disk rises rapidly as the disk is shocked. Radiation energy is roughly in equipartition with the kinetic energy at the peak. Then the ejecta expand, radiation energy decreases and converts to ejecta kinetic energy again via $pdV$ works. The ejecta is radiation pressure dominated, its internal energy is about three orders of magnitude smaller than kinetic and radiation energy.

At later time, the normalized kinetic energy increases to unity, suggesting the kinetic energy is comparable to the estimated $E_{\rm ej}$ (Equation~\ref{eq:Eej}), The three simulations show similar normalized energy evolution, indicating the energy conversion is insensitive to the tested $v_{*}$. The time delay between the peak of radiation energy density in the lower panel and the peak of luminosity in the upper panel is due to the combination of photon diffusion time through the ejecta and a small contribution from the time it takes the ejecta to expand out of the disk. The light travel time from the ejecta photosphere to the simulation domain boundary is negligible. Their proportionality to $v_{*}^{2}$ imply the adiabatic loss is also insensitive to the tested incident velocity. We measure the ejecta mass $M_{\rm ej}$ as described in Section~\ref{subsec:restult_densdisk} and find they are comparable in the three simulations due to the same $\Sigma_{\rm disk}$. 

The diffusion timescale $t_{\rm diff}\propto v_{\rm ej}^{-1/2}$ (Equation~\ref{eq:ldiff}) suggests that higher $v_{\rm ej}\sim v_{*}$ leads to shorter $t_{\rm diff}$. This is reflected in the slightly faster rise and decay of the luminosity and radiation energy density in Figure~\ref{fig:energy_lum_vz}. A secondary effect on $t_{\rm diff}$ is related to the temperature dependency of $\kappar$ and $\kappap$. In the ejecta, the gas temperature $T_{\rm ej}\sim T_{\rm rad}$ is in equilibrium with radiation temperature. Figure~\ref{fig:energy_lum_vz} shows that radiation and kinetic energy are comparable at the peak, so higher $v_{*}$ implies higher $T_{\rm rad}^{4}V\propto v_{\rm ej}^{2}\propto v_{*}^{2}$, where $V$ is the volume of ejecta. Assuming the expansion is adiabatic, ejecta gas has $\gamma=4/3$, $T\propto V^{1-\gamma}$, yielding temperature $T\propto v_{*}^{(5-4\gamma)/(2-2\gamma)}$. 

Performing similar one-dimensional analysis as Figure~\ref{fig:linecut_ddisk}, we find the $T\propto v_{*}^{(5-4\gamma)/(2-2\gamma)}$ relation is accurate within $10\%$ deviation. The ejecta velocity at $\rthr$ is $v_{\rm ej}=0.13c,~0.18c,~0.27c$ for H1\_md\_v0.07c, H1\_md\_v0.1c, H1\_md\_v0.15c forward ejecta. They all show $\rthp>\rthr$, the temperature at $\rthp$ are $9.6\times10^{4}K,~1.2\times10^{5}K,~1.5\times10^{5}K$ respectively. So when decreasing $v_{*}$, $T_{\rm ej}\sim T_{\rm rad}$ decreases, leading to larger $\kappar$ and $\kappap$. Together with the slower $v_{\rm ej}$, $t_{\rm diff}$ can be prolonged. 

\subsection{Relative Size of Star and Disk}\label{subsec:restult_hdisk}
In previous sections, we fix the scale height of the disk to be $H_{\rm disk}=R_{*}$. Equation~\ref{eq:ldiff} suggest that the luminosity weakly depends on the scale height $H_{\rm disk}$. For a star colliding with an accretion disk, $H_{\rm disk}$ varies when changing the location of collision and depends on the disk geometry. In addition, the relative size of the star and disk can change if the star is ablated after multiple interaction \citep{linial2023emri+, yao2025star}. In H3\_md\_v0.1c and H0.3\_md\_v0.1c, we set the disk scale height to be $H_{\rm disk}=3R_{*}$ and $H_{\rm disk}=0.3R_{*}$ to explore the effectffect of $H_{\rm disk}$. We fix the surface density when changing $H_{\rm disk}$, corresponding to disk midplane density $\rho_{\rm disk}=7.87\times10^{-8}\rm g~cm^{-3}$ and $\rho_{\rm disk}=7.87\times10^{-7}\rm g~cm^{-3}$.

\begin{figure}
    \centering
    \includegraphics[width=\linewidth]{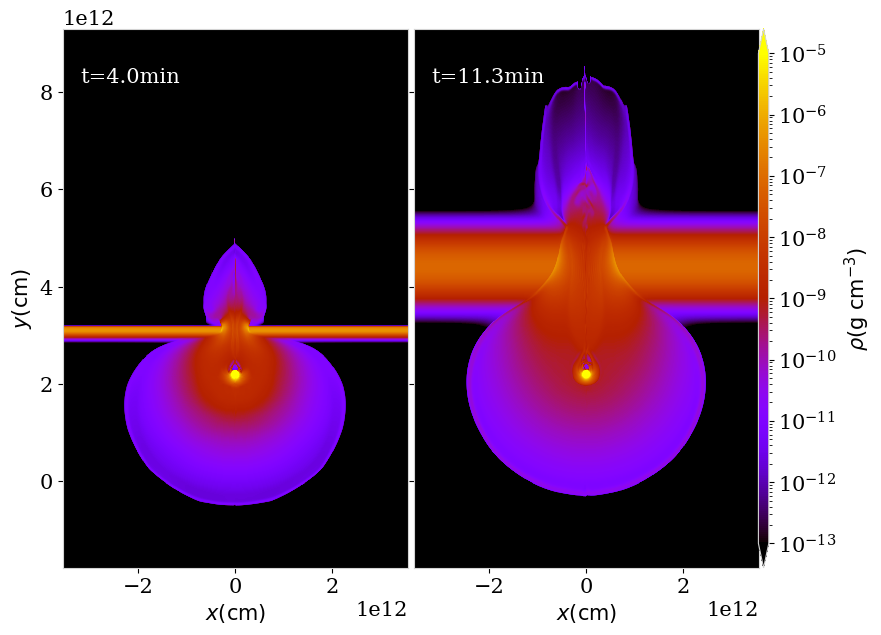}
    \caption{Gas density snapshot from H0.3\_md\_v0.1c (left) and H3\_md\_v0.1c (right). We show the x and y axis in unit of cm instead of normalize to $H_{\rm disk}$ due to different disk scale height. }
    \label{fig:hdisk_ej_dens}
\end{figure}

The ejecta morphology from the thin disk and the thick disk are different as shown in Figure~\ref{fig:hdisk_ej_dens}. In both cases, the forward and backward ejecta are asymmetric. With other parameters fixed, the run with $H_{\rm disk}=0.3R_{*}$ is similar to the run with $H_{\rm disk}=R_{*}$. But the run with $H_{\rm disk}=3R_{*}$ shows more extended backward ejecta. The ejecta's expanding speed is slower than the $H_{\rm disk}=0.3R_{*}$ and $H_{\rm disk}=R_{*}$ disks. 

\begin{figure}
    \centering
    \includegraphics[width=\linewidth]{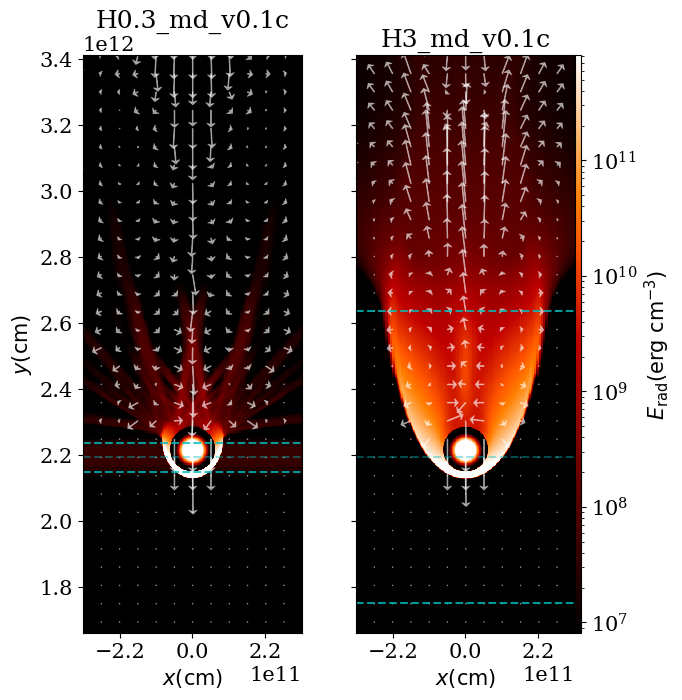}
    \caption{Gas radiation energy density snapshots from H0.3\_md\_v0.1c and  H3\_md\_v0.1c. The white arrows shows the normalized velocity field in the disk frame. In each snapshot, the three dashed cyan lines show the midplane of disk, and $\pm 2H_{\rm disk}$ from the midplane. In the left plot (H0.3\_md\_v0.1c), the beamed pattern above the disk is due to the angular discretization of radiation field in the numerical scheme in extreme optically thin region, which is referred to as ``ray effect''. }
    \label{fig:hdisk_shock}
\end{figure}

The bow shock structure is relatively insensitive to $H_{\rm disk}$. In Figure~\ref{fig:hdisk_shock}, the bow shock opening angles are similar in H3\_md\_v0.1c and H0.3\_md\_v0.1c, the radiation energy density at the shock front is also comparable. However, there are larger volume of optically thick gas in H3\_md\_v0.1c, leading to enhanced radiation energy density in the disk. The diffusion timescale in the disk is roughly $t_{\rm diff,disk}\sim\tau_{\rm disk}H_{\rm disk}/c$, so when fixing $\tau_{\rm disk}$, larger $H_{\rm disk}$ suggests slower diffusion of photons that are produced at the shockfront. The shocked gas also remains in the disk for longer time, energy loss in the disk leads to slower expansion. The forward and backward ejecta are more symmetric in size. 

\begin{figure}
    \centering
    \includegraphics[width=\linewidth]{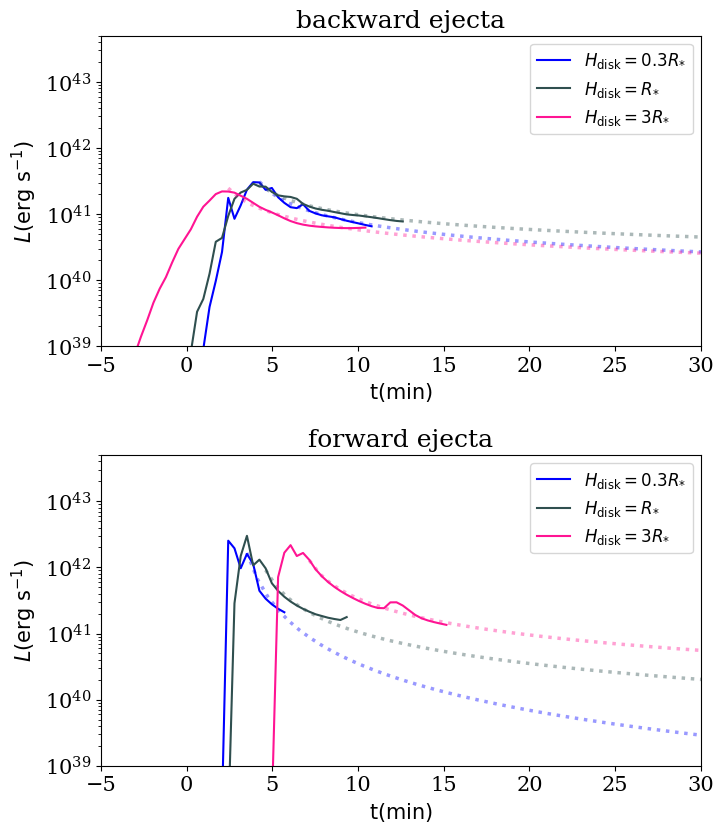}
    \caption{Estimated luminosity of backward ejecta (upper panel) and forward ejecta (lower panel). The blue, black, pink lines correspond to H0.3\_md\_0.1c ($H_{\rm disk}=0.3R_{*}$), H1\_md\_0.1c ($H_{\rm disk}=R_{*}$) and H3\_md\_0.1c ($H_{\rm disk}=3R_{*}$). The dotted lines show power law fittings (Equation~\ref{eq:expdecay}). The time axis is the time since star is at the disk midplane.}
    \label{fig:lum_hdisk}
\end{figure}

Figure~\ref{fig:lum_hdisk} shows that the estimated total luminosity of H3\_md\_v0.1c and H0.3\_md\_v0.1c. The run with $H_{\rm disk}=0.3R_{*}$ disk is similar to the fiducial run with $H_{\rm disk}=R_{*}$. However, the run with thicker disk $H_{\rm disk}=3R_{*}$ shows a prolonged light curve as result of longer disk diffusion time and slower evolving ejecta. The best fitting parameters for the power law backward ejecta decay indicates slope $\beta=-0.90,~-1.12$, and timescale $\sigma=1.46,~2.62$min for H0.3\_md\_v0.1c and H3\_md\_v0.1c. For the forward ejecta, the best fitting parameters are slope $\beta=-0.68,0.86$ and $\sigma=0.72,~0.74$ min for H0.3\_md\_v0.1c and H3\_md\_v0.1c. 

\subsection{Opacity Structure of Ejecta}\label{subsec:result_Opacity}
The ejecta $t_{\rm diff}$ affects the light curve decay timescale. When comparing the ejecta cooling emission to QPE flares, it can affect the QPE flare duration. Equation~\ref{eq:ldiff} estimates $t_{\rm diff}$ by assuming photons diffuse out of expanding ejecta when the diffusion speed $c/\tau$ is comparable to the ejecta speed $v_{\rm ej}$. For quantities in Equation~\ref{eq:ldiff}, the estimations of $M_{\rm ej}$ and $v_{\rm ej}$ are broadly consistent with simulation results. In the simulations, however, the ejecta opacity structure can be more complex than scatter dominated opacity. The absorption and emission opacity influence the energy exchange between radiation and gas. In this section, we discuss the opacity structure of ejecta in the simulations.  

\textit{Free-Free opacity}: For the ejecta profiles shown in Figure~\ref{fig:linecut_ddisk}, the typical gas temperature at $\taueffp=c/v$ is roughly $\sim 1-2\times10^{5}$K, which is lower than the blackbody temperature indicated by observed QPE emission in soft X-ray. Using $\taueffr=c/v$ yields slightly higher gas temperature, but it is typically below $10^{6}$K. 

One theoretical resolution relates to the inefficient photon production via free-free emission \citep{nakar2010early,linial2023emri+,vurm2025radiation}. Before the ejecta significantly expand and still contain high energy photons, over a timescale that is much shorter than the dynamical time, approximately radiation does not exchange energy with gas, so the total photon energy is conserved. When the photon energy distribution evolves towards blackbody spectrum, the photon number must increase so that the average energy per photon decease. Effectively, efficient photon production is essential to reach blackbody spectral energy distribution (SED). The photon production rate by free-free emission can be estimated as $\dot{n}_{
\rm ff}(\rho,T)\propto\rho^{2}T^{-1/2}$ \citep{nakar2010early}. If ejecta density is sufficiently low, free-free emission might not be able to produce enough photons to establish a blackbody spectrum. The photon deficit yields a spectrum with higher photon energy compared to a blackbody spectrum.

\begin{figure}
    \centering
    \includegraphics[width=\linewidth]{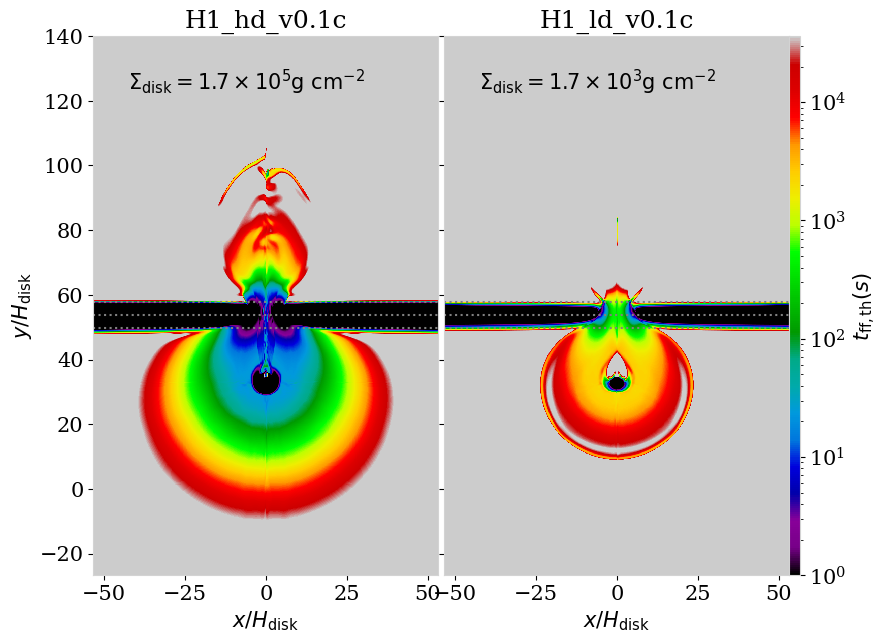}
    \caption{Estimated timescale $t_{\rm ff,th}$, which is the timescale to produce enough photons to establish a blackbody SED by free-free emission. Left is for the high surface density disk H1\_hd\_v0.1c and right is for the low surface density disk H1\_ld\_v0.1c, the snapshot time is $t=5.0$ min since the star is at disk midplane, corresponding to the third column in Figure~\ref{fig:result_density_h1}. }
    \label{fig:tffth_ddisk}
\end{figure}

The timescale to produce enough photons to establish a blackbody SED via free-free emission, which we refer as  free-free photon production time, can be estimated as $t_{\rm ff,th}=n_{\rm BB}/\dot{n}_{\rm ff}$, where $n_{\rm BB}\approx aT^{4}/3k_{\rm B}T$ approximates the photon number in blackbody SED \citep{nakar2010early,linial2023emri+}. Figure~\ref{fig:tffth_ddisk} shows $t_{\rm ff,th}$ for H1\_hd\_v0.1c and H1\_ld\_v0.1c, with the highest and lowest disk surface density. 

The ejecta temperature in H1\_hd\_v0.1c is higher than  H1\_ld\_v0.1c up to one order of magnitude, while the ejecta density is higher than H1\_ld\_v0.1c about two orders of magnitude, roughly proportional to $\Sigma_{\rm disk}$. The backward ejecta in both simulations shows $t_{\rm ff,th}\gtrsim 30$min, smaller $t_{\rm ff,th}\lesssim 10$min are only achieved near the base of the ejecta, where the density is higher. The forward ejecta in  H1\_ld\_v0.1c has $t_{\rm ff,th}\gtrsim 30$min except for the base, while forward ejecta in  H1\_hd\_v0.1c includes regions with relatively short $t_{\rm ff,th}\lesssim 10$min. The free-free photon production time $t_{\rm ff, th}\propto\rho^{2}T^{-1/2}$. The ejecta density is roughly proportional to disk density $\rho_{\rm disk}\propto\Sigma_{\rm disk}$. Given the most region of ejecta are still optically thick to absorption, the temperature of ejecta roughly $T_{\rm ej}\propto \rho_{\rm ej}^{1/3}\propto\Sigma_{\rm disk}^{1/3}$. So roughly $t_{\rm ff,th}\propto\Sigma_{\rm disk}^{13/6}$ between these two simulations. Lower surface density disk such as H1\_ld\_v0.1c is more likely to show longer $t_{\rm ff,th}$ and is more favorable to produce hardened spectrum due to inefficient photon production by free-free emission.

\textit{Atomic opacity (bound-free)}: The above estimation of $t_{\rm ff,th}$ assumes the free-free opacity is the main source of emissivity. In the simulations, however, the photo-ionization (bound-free) opacity contribute to the Planck mean absorption opacity in the ejecta, as informed by TOPs opacity table assuming solar abundance. In the outer ejecta where temperature $T\lesssim10^{6}$K, we have $\kappap$ and $\kappar$ larger than free-free opacity, and $\kappap\gg\kappar\sim\kappas=0.32 \rm g^{-1}cm^{2}$. The additional bound-free opacity can enhance the energy exchange between radiation and gas, potentially enhancing photon production.

Another potential impact is the bound-free absorption opacity may lead to longer diffusion time than $t_{\rm diff}$ estimated with only $\kappa_{\rm es}$ (Equation~\ref{eq:ldiff}). We estimate an effective diffusion time at radius $r$ in the ejecta by:
\begin{equation}\label{eq:tdiff_eff}
    t_{\rm diff,eff}=\taueffr (R_{\rm out}-r)/c,
\end{equation}

\begin{figure}
    \centering
    \includegraphics[width=\linewidth]{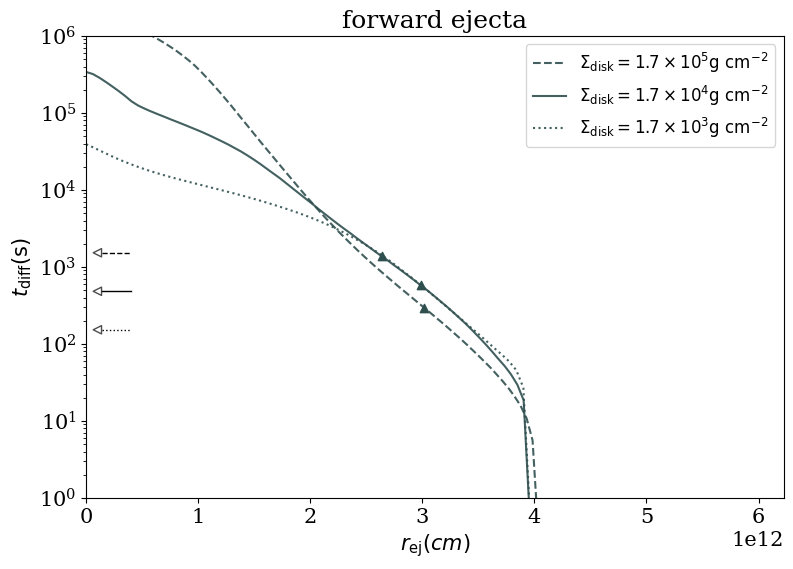}
    \caption{Effective diffusion timescale (Equation~\ref{eq:tdiff_eff}) along the sampling line similar to Figure~\ref{fig:linecut_ddisk} in the forward ejecta for H1\_hd\_0.1c ($\Sigma_{\rm disk}=1.7\times10^{5}\rm g~cm^{-2}$, dashed lines), H1\_md\_0.1c ($\Sigma_{\rm disk}=1.7\times10^{4}\rm g~cm^{-2}$, solid lines), H1\_ld\_0.1c ($\Sigma_{\rm disk}=1.7\times10^{3}\rm g~cm^{-2}$, dotted lines). The triangles label the location of $\taueffr=c/v$, $t_{\rm diff}=2.97\times10^{2}\rm s,~5.78\times10^{3}s,~1.4\times10^{3}s$ in H1\_hd\_v0.1c, H1\_md\_v0.1c, H1\_ld\_v0.1c. The three triangles next to the left y axis labels are the estimated diffusion time $t_{\rm diff}$ with $\kappa_{\rm es}$ (Equation~\ref{eq:ldiff}), from higher to lower correspond to descending $\Sigma_{\rm disk}$. }
    \label{fig:tdiff_ddisk}
\end{figure}

Figure~\ref{fig:tdiff_ddisk} shows the estimated effective diffusion time for H1\_hd\_v0.1c, H1\_md\_v0.1c and H1\_ld\_v0.1c using the same sampling line for the forward ejecta as in Figure~\ref{fig:linecut_ddisk}. We label the location where $\taueffr=c/v$ by triangles, $t_{\rm diff}=2.97\times10^{2}\rm s,~5.78\times10^{3}s,~1.4\times10^{3}s$ for H1\_hd\_v0.1c, H1\_md\_v0.1c, H1\_ld\_v0.1c. Compare to Equation~\ref{eq:tdiff}, which gives $t_{\rm diff}=1.53\times10^{3}\rm s,~4.83\times10^{2}s,~1.53\times10^{2}s$.

Comparing the three simulations, higher $\Sigma_{\rm disk}$ yields smaller $t_{\rm diff, R}$. This contradicts the dependency of $t_{\rm diff}\propto\Sigma_{\rm disk}^{1/2}$ (Equation~\ref{eq:tdiff}), which arises from the fact that higher $\Sigma_{\rm disk}$ leads to larger $\rho_{\rm ej}$ and thus optical depth. However, the $T_{\rm ej}$ is also higher when increasing $\Sigma_{\rm disk}$, so the opacity is reduced due to less bound-free contribution. A relevant implication is that if interpreting the duration of QPE flares as the diffusion time of ejecta, adopting only electron scatter opacity can overestimate the ejecta mass \cite[][see also recent observation]{chakraborty2025discovery, arcodia2025srg}. The bound-free opacity effect could potentially extend parameter space for low density disks that were previously excluded due to too short diffusion time. 

However, the bound-free opacity can be suppressed if the ionization fraction of ejecta is high. For example, \citet{nayakshin2004x} showed that the opacity for photon energy above 1KeV can be reduced by orders of magnitude when the ionization parameter is $\xi\gtrsim10^{2}$. Recent spectral modeling \citet{chakraborty2025discovery} estimates ionization parameter $\xi\gtrsim10^{4}$ in the ejecta cloud. The soft X-ray photons from the shocked disk are typically below 1KeV, but over-ionization can also potentially affect the bound-free opacity in this photon energy range, which needs to be investigated in follow-up work.

\section{SED evolution: Multi-group RHD Simulations}\label{sec:result_multi}
In this section, we discuss results from multi-group radiation hydrodynamic simulations, including six \textit{dynamic runs} that evolve the star-disk collision dynamically with multi-group RHD including 10 photon frequency groups. In addition, we explore Compton scattering's effect with two \textit{post-process runs} that only evolve the radiation field to steady state according to the thermal property obtained from \textit{dynamic runs}, without evolving hydrodynamics.

\subsection{Multi-group Opacity}\label{subsec:multgroup_opacity}
\textit{Frequency-dependent Opacity}
We adopt multi-group opacity from TOPS Opacity database \citep{colgan2016new}. The TOPS table returns Rosseland mean $\kappar$ and Planck mean opacity $\kappap$, which are averaged within each photon frequency group. The Rosseland mean opacity is in the format of the sum of Rosseland mean opacity and electron scatter opacity $\kappar+\kappa_{\rm es}$.

We retrieve multi-group opacities for 10 photon frequency groups ranging from $h\nu=10^{-3}$KeV to $h\nu=4$KeV. This corresponds to webpage entries of 11 photon frequencies with lower and upper bound of $10^{-3}$KeV and $10$KeV. The first group represents $0$KeV to $10^{-3}$KeV, the last group represents $4$KeV to infinity. We request 100 density points from $10^{-17}-10^{-6}\rm g~cm^{-3}$ , 69 default temperature points from $5\times10^{-4}$KeV to $10$KeV. 

\begin{figure}
    \centering
    \includegraphics[width=\linewidth]{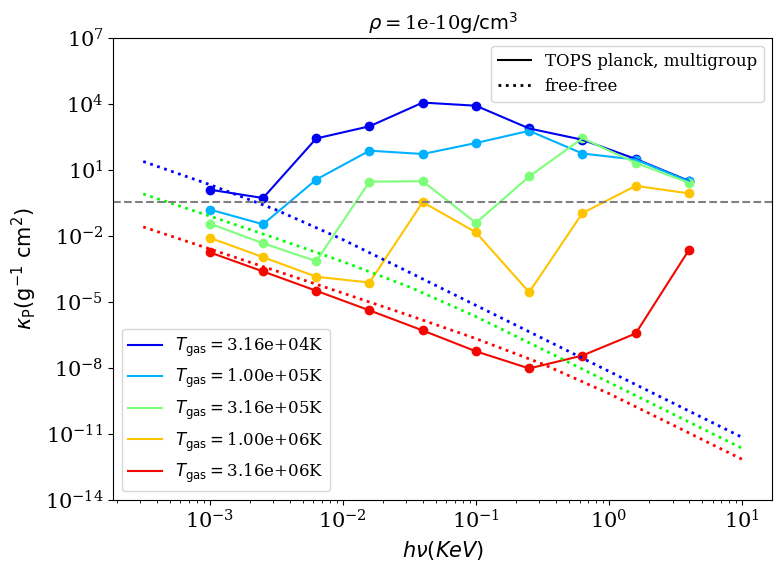}
    \caption{The multi-group Planck mean opacity from TOPS \citep{colgan2016new} for fixed density $\rho=10^{-10}\rm g~cm^{-3}$ (solid lines) with solar abundance. This retrieval includes 10 photon frequency groups that are logarithmically spaced between $10^{-3}$KeV to $4$KeV (marked by dots). Each line corresponds to a specific gas temperature: $3.16\times10^{4}$K (deep blue), $10^{5}$K (light blue), $3.16\times10^{5}$K (green), $10^{6}$K (orange) and $3.16\times10^{6}$K (red). The dotted lines with the same color correspond to free-free opacity for the same gas temperature. The gray dashed line labels $\kappas=0.34\rm g^{-1}~cm^{2}$.}
    \label{fig:kappap_tops}
\end{figure}

We assume standard solar abundance X=0.735, Y=0.248, Z=0.017 \citep{grevesse1998standard} for the mixture. In the simulations, the opacity of each cell is linearly interpolated in temperature and density grid of TOPs data. For low density and high-temperature regions where TOPS does not provide multi-group opacity data (assuming local thermal equilibrium (LTE) is not well satisfied), we replace the absorption opacity by free-free opacity. Such ad-hoc treatment can be improved by integrating non-LTE calculations more consistently in the future works.

Figure~\ref{fig:kappap_tops} shows a subset of the tabulated Planck mean opacity for a fixed gas density $\rho=10^{-10}\rm g~cm^{-3}$, including five representative temperatures ranging from  $3.16\times10^{4}-3.16\times10^{6}$K as the colored solid lines. The Rosseland mean opacity for the same density, temperature and photon energy groups are smaller than Planck mean opacity by orders of magnitudes. It becomes $\kappar\lesssim\kappas$ when $T>6.3\times10^{5}$K for all shown photon energy groups. 

As a comparison, the dotted lines show approximated free-free opacity. For sufficiently high gas temperature $3.16\times10^{6}$K, $\kappa_{\rm ff}\approx\kappap$ well. For temperature groups between $10^{4}-10^{6}$K, $\kappa_{\rm ff}<\kappap$. For example, the tabulated opacity at $T=3.16\times10^{5}$K is larger than electron scatter for photon frequency groups of $\nu=1.58\times10^{-2}$KeV, $3.98\times10^{-2}$KeV and $\geq2.51\times10^{-1}$KeV.  The opacity contribution from bound-free (ionization) is consistent with the trend observed in the frequency-integrated $\kappap$ and $\kappar$ in previous sections, where the gray opacities are also larger than free-free opacity due to the bound-free processes. The enhanced opacity might facilitate the thermalization of the photon spectrum.

For these temperatures relevant to the ejecta, an interesting trend is the slight drop of $\kappap$ near the photon frequency of $\nu\approx0.1-0.3$KeV. The trend is also observed in other similar gas densities $\rho\approx10^{-12}-10^{-8}\rm g~cm^{-3}$ that are representative for the ejecta. This multi-group opacity effect may help to reduce the optical depth for photons with energy of a few $100$eV, which are relevant to the soft X-ray QPE flares. 

In the \textit{dynamic runs}, we first run gray RHD simulation when the star just arrives at disk midplane, before backward ejecta forms. Then we map the gray RHD simulation as the initial condition for multi-group simulations. The mapping utilizes the density, velocity, gas and radiation temperature information from the gray simulation, it assumes a blackbody spectrum at $T_{\rm rad}$ and assigns energy density to each photon frequency groups while assuming radiation intensities are isotropic in all angular directions \citep{jiang2022multigroup}. These multigroup simulations then evolve full hydrodynamics and radiation transfer. 

However, we neglect two processes in these \textit{dynamic runs}. First, we drop the frequency group shift (Doppler shift) during frame transformation between the co-moving frame and the disk frame, which may be secondary given the typical ejecta velocity $v_{\rm ej}<0.2c$. Second, we do not include the thermal Compton effect due to relatively low gas temperature in the ejecta $T_{\rm ej}\lesssim5\times10^{6}$K. 

We explore potential effect from Compton scattering by \textit{post-processing} simulations, which includes both processes neglected in \textit{dynamic runs}: bulk and thermal Compton. In these simulations, we restart the \textit{dynamic runs} from multiple time snapshots, but turn off hydrodynamic evolution and allowing only radiation field to evolve. The radiation field will converge to quasi steady-state, but the coupling and feedback between radiation and gas will not be captured. 

\subsection{Fiducial Multigroup Collision: H1\_md\_0.1c}\label{subsec:multigroup_fiducial}
\subsubsection{Effect of Multi-group Opacity on SED}
We map H1\_md\_0.1c at $t=7.27$min as the initial condition for its \textit{dynamic} multi-group run. The evolution of hydrodynamical variables such as gas density, temperature and velocity are roughly consistent with the gray RHD results, with moderate ejecta morphological differences. The total radiation energy in the domain summing over all frequency groups also matches with the gray total radiation energy density. 

Similar to gray RHD run, we estimate the luminosity by integrating the radiation flux over the simulation domain boundary: $L_{\nu}=\int \textbf{F}_{\nu}\cdot \textbf{dA}$ for each photon energy group $\nu$. We make sure that $L_{\nu}$ is measured outside the $\rthp$ and $\rthr$ before the ejecta expands beyond the simulation domain.

\begin{figure}
    \centering
    \includegraphics[width=\linewidth]{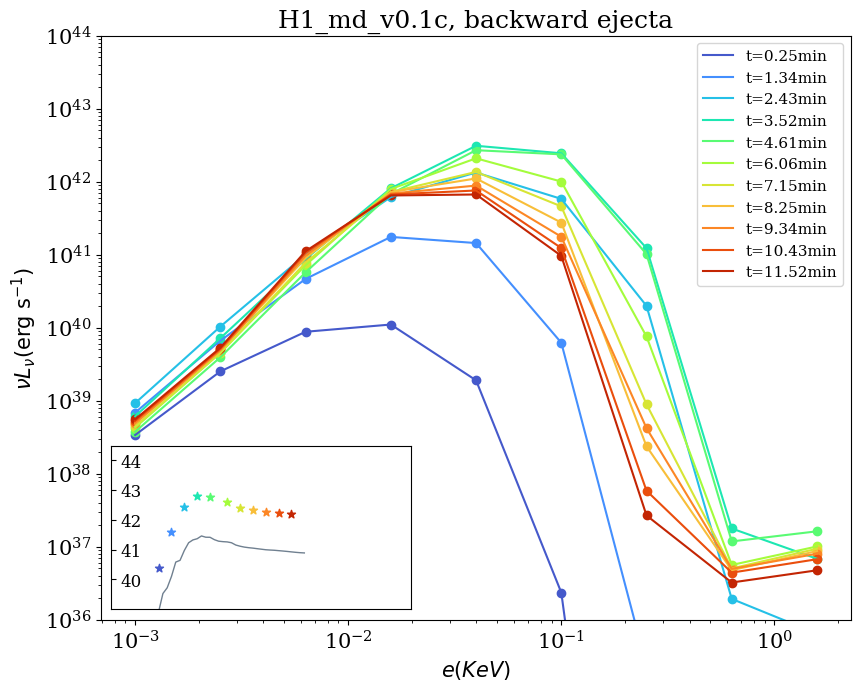}
    \includegraphics[width=\linewidth]{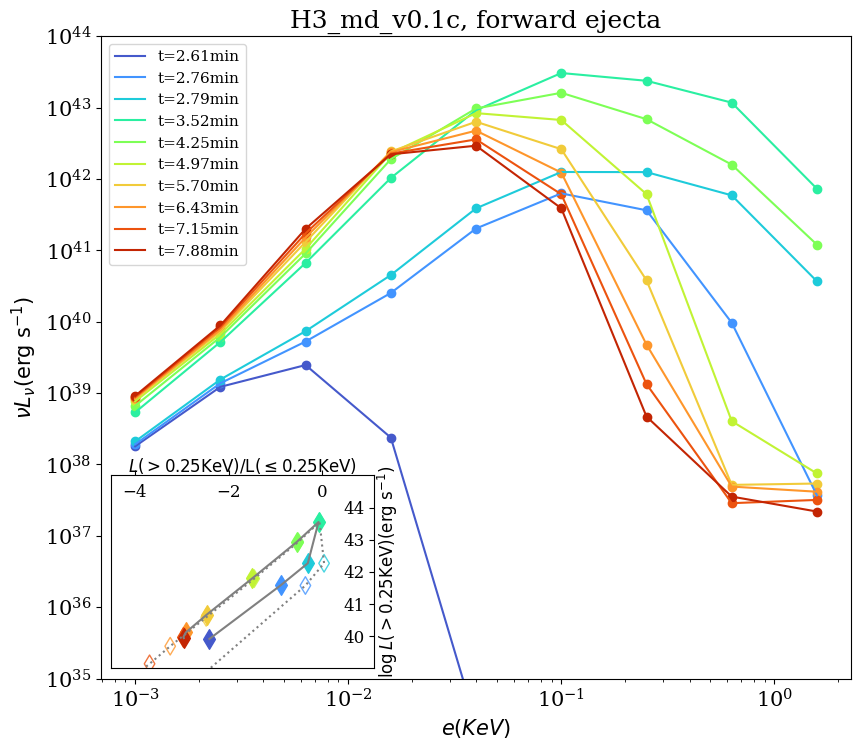}
    \caption{Snapshots of estimated broad-band SED from H1\_md\_v0.1c multi-group \textit{dynamic} run. We drop the last group of $4$KeV to infinity. The time label is the time since the star is at the disk midplane. The first plot (upper panel) is the backward ejecta. In the inset plot, we plot the shape of gray light curve as the solid gray line, the y-axis is log-spaced and labeled, the x-axis is linear-spaced and not labeled. The star marks correspond to the sum of $\nu L_{\nu}$ of all frequency groups from multi-group simulation, i.e. an estimation of bolometric luminosity from multi-group run. They are higher than the gray RHD runs values due to the frequency-dependent opacity. The second plot (lower panel) is the forward ejecta. The inset figure shows its ``hardness-luminosity'' cycle. We defined a ``soft X-ray to UV-optical'' ratio as the ratio of luminosity above and below 250eV, which is the x-axis. The y-axis shows luminosity above 250eV. The solid line and filled data points include emission from a standard, thin accretion disk, the dashed line and hollow data points are directly from simulation.}
    \label{fig:multi_spec_h03}
\end{figure}

To approximate the broad-band SED, we show multiple snapshots of $\nu L_{\nu}$ as function of $h\nu$ in Figure~\ref{fig:multi_spec_h03}. We adopt the left edge value of each frequency group for $h\nu$ when calculating the luminosity, i.e. $\nu L_{\nu}=\nu_{i}(\int_{\nu_{i}}^{\nu_{i+1}}L_{\nu}d\nu/\int_{\nu_{i}}^{\nu_{i+1}}d\nu)$, the x-axis represents $h\nu=h\nu_{i}$. 

The backward ejecta luminosity rises smoothly, from $t=0.2$ min since the star is at disk midplane, the SED energy increases to peak in $\sim4.4$ minutes. The decay is slower than the rise, which decreases for $\sim7$ minutes from the peak until the ejecta expands beyond the simulation domain. The spectral energy peaks $40-100$eV, with luminosity at the order of $\nu L_{\nu}\sim 10^{42}\rm erg~s^{-1}$. The lower energy bands $h\nu\lesssim20$eV can be approximated by a black-body spectrum corresponding to temperature $\sim 10^{5}$K, consistent with the gas temperature near $\taueffp$. The higher energy bands appear as an excess in addition to the blackbody component. The SED sharply cut off about $1$KeV, the ninth photon frequency group $1.58$KeV is affected by the low spectral resolution that close to the last group. 

The forward ejecta SEDs show rapid rise due to the breakout emission when radiation at the shock front diffuses out from the disk. For example, the SEDs at $t=2-3$ min roughly represent this breakout emission. The bolometric luminosity reaches to $\nu L_{\nu}\sim10^{43-44}\rm erg~s^{-1}$, and peaks in the spectral energy $h\nu\sim250$eV, with a high energy tail extending beyond $1$KeV. However, the breakout signal lasts briefly, about 2 minutes in H1\_md\_v0.1c due to the small amount of mass the bow shock sweeps through.

Followed by the breakout signal, the decay is significantly slower than the rise, which lasts for 5 minutes before the ejecta expands beyond from the simulation domain. In the decay, the photon energy peak quickly shifts from $\sim100$eV to $\sim30$eV. Similar to the backward ejecta, the lower energy SED $h\nu\lesssim20$eV can be approximated by a black-body component corresponding to temperature $\sim 10^{5}$K. The luminosity for $h\nu\gtrsim630$eV is below $10^{40}\rm erg~s^{-1}$ except near the breakout emission. 

In the inset figure, we show the ``hardness-luminosity'' evolution of these SEDs. We define a ``soft X-ray to UV-optical ratio'' as the ratio between total luminosity above and below 250eV: $L(>0.25\rm KeV)$/$L(\leq0.25 \rm KeV)$, and a ``soft X-ray luminosity'' as the total luminosity above 250eV: $L(>0.25\rm KeV)$. The dashed line and hollow data points are directly from simulations, derived from the SEDs that are plotted in the main figure. 

In addition, we show an estimated ``hardness-luminosity'' evolution including the analytical spectrum of a quiescent disk. The solid line and filled data points include a thin, thermal disk for $M_{\rm BH}=10^{6}M_{\odot}$. We assume the disk accretion rate $\dot{M}=0.1\dot{M}_{\rm Edd}$, with inner radius $r_{\rm in}=6r_{\rm g}$ and outer radius $r_{\rm in}=1000r_{\rm g}$. For the disk spectrum, each radial annulus is assumed to be emitting at local blackbody temperature and then integrated between the inner and outer radius to infer the disk spectrum.

The ``hardness-luminosity'' evolution show the trend that as the luminosity rises, SED becomes harder. For example, when the $L(>0.25\rm KeV)$ rises from $\sim 10^{41}\rm erg~s^{-1}$ to  $\sim 10^{44}\rm erg~s^{-1}$, the ratio $L(>0.25\rm KeV)$/$L(\leq0.25 \rm KeV)$ increases three orders of magnitude. During the luminosity decay, the ``hardness'' ratio cycles back to $L(>0.25\rm KeV)$/$L(\leq0.25 \rm KeV)\sim10^{-3}$. This is roughly consistent with the hysteresis cycle that often seen in QPE flares, where the flare effective temperature increases a few times during the rise and decreases to nearly the initial temperature during the decay. The ``hardness ratio-luminosity'' variation in our simulation shows higher amplitudes compared to observed QPE flare variations \cite[e.g.][]{arcodia2022ero1}, but the discrepancy can be related to definitions of hardness ratio and luminosity. However, the timescale of the hysteresis cycle in the simulations is less than 10 minutes, which is significantly briefer than the most observed cycles \citep[][e.g]{arcodia2021x,chakraborty2024testing,nicholl2024quasi}. We discuss potential directions to prolong the simulated soft X-ray flare and alternative scenarios in Section~\ref{subsec:favored_condition}.

\begin{figure}
    \centering
    \includegraphics[width=\linewidth]{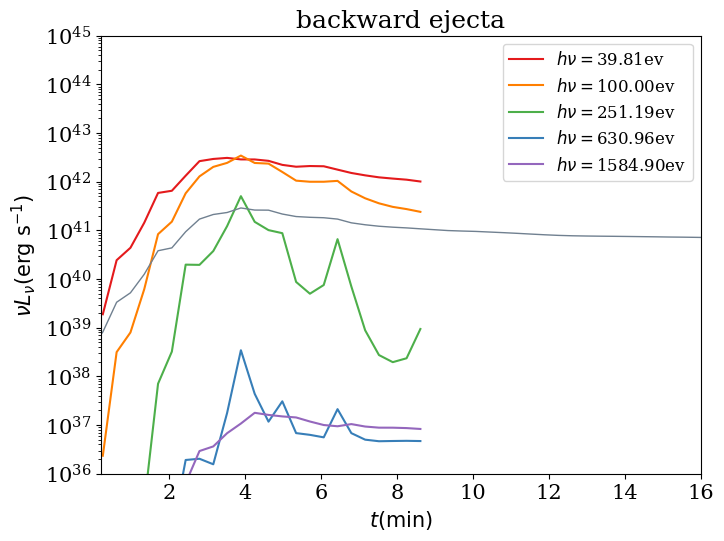}
    \includegraphics[width=\linewidth]{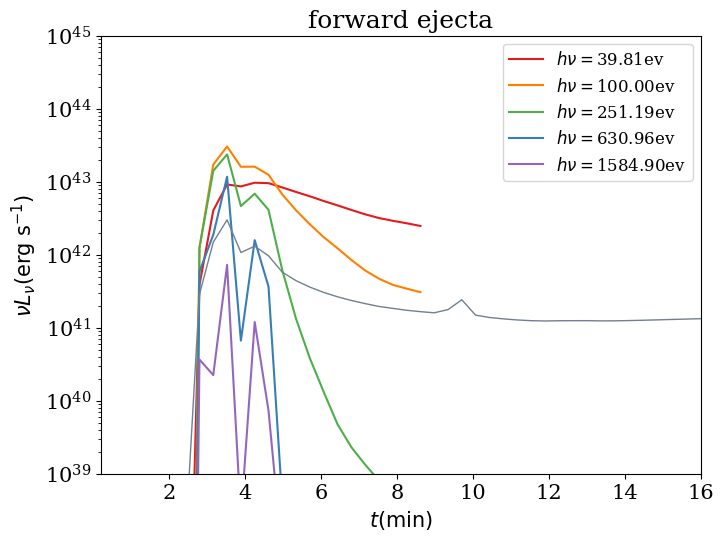}
    \caption{Evolution of $\nu L_{\nu}$ of five photon frequency groups: $40$eV (red), $100$eV (orange), $250$eV (green), $630$eV (blue) and $1.58$KeV (purple) for the backward ejecta (upper) and forward ejecta (lower). The time axis is the time since the star is at the disk midplane. The gray solid line in each plot shows the bolometric luminosity measured in gray RHD simulation, which can underestimate the bolometric luminosity compared to the multi-group simulation.}
    \label{fig:multi_lc_h03}
\end{figure}

A significant difference between the multi-group simulations and the gray simulations is the total luminosity. The inset plots in the upper panel of Figure~\ref{fig:multi_spec_h03} show that the total luminosity summed over all frequency groups in the multi-group runs shares the same trend as the gray RHD luminosity, but can be higher than the gray RHD luminosity for about one order of magnitude. We also find the same trend in the forward ejecta. In Figure~\ref{fig:multi_lc_h03}, we show the evolution of $\nu L_{\nu}=\nu_{i}\int_{\nu_{i}}^{\nu_{i+1}}L_{\nu}d\nu/(\nu_{i+1}-\nu_{i})$ for five representative groups together with the bolometric luminosity from gray RHD counterpart, effectively $\nu L_{\nu}=\int_{0}^{\infty}L_{\nu}d\nu$. The time sample separation is 20 seconds. 

For the groups that are related to the soft X-ray flares, $100$eV, $251$eV and $631$eV, the luminosities is $\gtrsim10^{42}\rm erg~s^{-1}$ near the peak and in the decay. The higher multi-group luminosity compared to the gray luminosity demonstrates an important multi-group opacity effect: the photosphere temperature for each frequency group can differ from the temperature at the gray photosphere. When comparing the gray and multi-group runs, the total radiation energy produced during the star-disk collision is comparable, suggesting the conversion from kinetic energy to radiation energy by the bow shock is roughly unaltered. However, the multi-group run has more radiation flux diffusing out from ejecta compared to the gray run, increasing the luminosity from those photon frequency groups with smaller optical depth than the frequency-averaged optical depth. 

\subsubsection{Effect of Compton scattering on SED}\label{subsection:multi_compton}

In this section, we explore the potential effect of Compton scattering by \textit{post-processing} runs, and we also show results of \textit{post-processing} without Compton scattering as comparisons. In these simulations, we evolve the radiation field until it reaches quasi steady-state while fixing gas density, velocity and temperature distributions. We summarize the resulting SEDs as a blackbody component peaks optical-UV band and potential Comptonized tail extends to energies above KeV. 

Note that the approach of fixing hydrodynamics in these \textit{post-processing} runs are equivalent to assuming that the timescale for radiation field to reach steady-state is significantly shorter than the hydrodynamic timescale, so the hydrodynamic flow does not change significantly, and thus can be approximated as static. This is, however, not likely to be well-satisfied during the star-disk collision due to the high incident velocity and short hydrodynamic timescale. The effect of Compton scattering including a dynamical feedback between radiation and gas will be explored in future work, and we speculate their SEDs shape may fall in between the \textit{dynamic} SEDs and \textit{post-processing} SEDs discussed in this section.

\begin{figure}
    \centering
    \includegraphics[width=\linewidth]{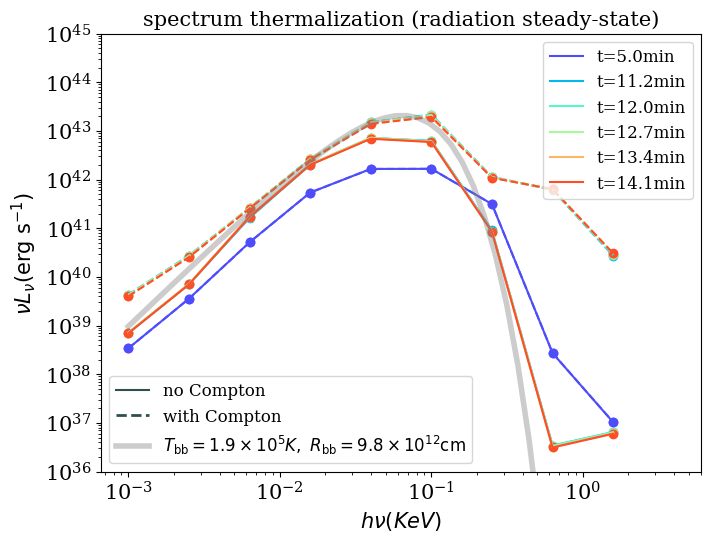}
    \caption{The SED snapshots from H1\_ld\_v0.1c backward ejecta in the \textit{post-processing} run restarting from t=5.0min after the star is at disk midplane. The blue line is from the beginning of \textit{post-processing} run, equivalent to the \textit{dynamic} run results. The other color lines show series of SEDs when radiation evolves to quasi-steady state, so they are similar. The solid lines shows \textit{dynamic} run without Compton scattering, the dashed lines shows \textit{dynamic} run with Compton scattering. The thick gray solid line shows a single blackbody spectrum at $T=1.9\times10^{5}$K for reference. The time label is the time since the star is at disk midplane.}
    \label{fig:multi_compton_evolve}
\end{figure}

First, to illustrate the process of radiation field relax to quasi-steady state, Figure~\ref{fig:multi_compton_evolve} shows the SED at the beginning of \textit{post-processing} run H1\_ld\_0.1c (blue line) and the quasi-steady state SEDs (other lines). The initial SED (blue solid line) from \textit{dynamic} run cannot be well approximated by a single blackbody spectrum. Starting from t=11.2min, the SEDs are not significantly evolving, so the radiation field relaxes to quasi-steady state after about six minutes. Without Compton scattering (the solid lines), the non-blackbody SED at t=5min is reprocessed towards a single blackbody spectrum. With Compton scattering (the dashed lines), the energy groups below $100$eV can still be approximated as blackbody.  

We plot a single blackbody spectrum $L_{\rm BB}=4\pi R_{\rm BB}^{2}\times\frac
{2h}{c^{2}\nu^{3}(\rm exp(h\nu/k_{B}T_{ BB})-1)}$ in Figure\ref{fig:multi_compton_evolve} for reference, with $\nu$ in Hz and $T_{\rm BB}$ in K. For groups $h\nu\gtrsim100.0$eV, the soft X-ray luminosity is enhanced to $\nu L_{\nu}\sim10^{42-43}\rm erg~s^{-1}$. The spectral energy for groups $h\nu\lesssim39.8$eV also slightly increases. Interesting, we find that the Compton scattering is primarily from bulk Compton instead of thermal Compton, where the gas is compressed by the shock in the disk, introducing strong velocity divergence for shock-heated gas with $v\gtrsim0.1c$ . We test turning off thermal Compton and find negligible differences. We note that the frequency resolution in the multi-group simulations is insufficient to prefer or disfavor spectrum common fitting models in the soft X-ray band \cite[e.g.][]{chakraborty2024testing, pasham2024alive}. 

\begin{figure}
    \centering
    \includegraphics[width=\linewidth]{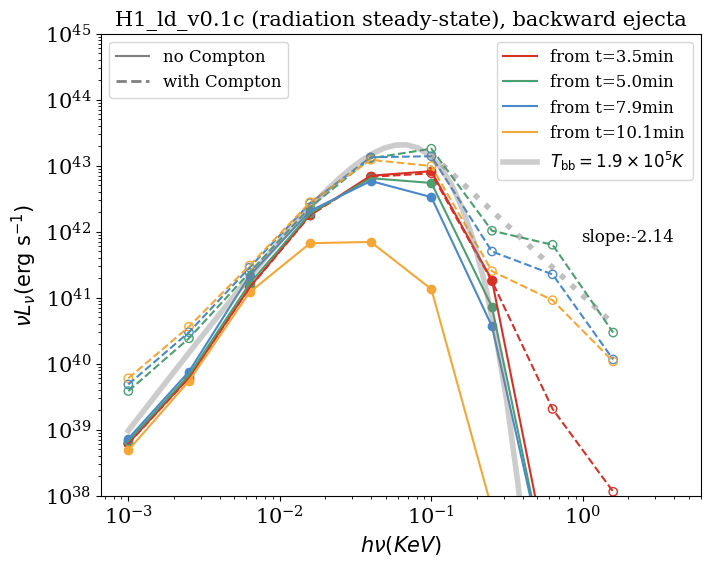}
    \includegraphics[width=\linewidth]{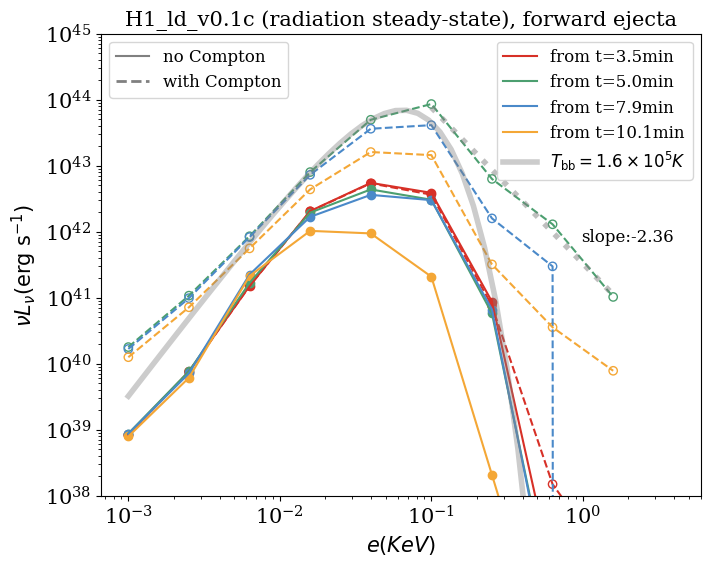}
    \caption{Quasi-steady state SEDs in \textit{post-process} run for H1\_ld\_v0.1c. The first plot is for the backward ejecta, and the second plot is for the forward ejecta. In each plot, the solid lines and filled circles are quasi-steady state SEDs without Compton scattering, the dashed lines and hollowed circles are quasi-steady state SEDs with Compton scattering. The four sets of dashed and solid lines are with colors corresponding to four restarting times since the star is at disk midplane: t=3.5min (red), t=5.0min (green), t=7.9min (blue) and t=10.1min(yellow). The gray solid line shows single temperature blackbody spectrum for reference, the dotted gray line shows a fitting of Compton tail with slope labeled by the text.}
    \label{fig:multi_compton}
\end{figure}

Figure~\ref{fig:multi_compton} shows the quasi-steady state SEDs that are evolved from four distinct times in H1\_ld\_v0.1c: t=3.5min, 5.0min, 7.9min and 10.1min since the star is at disk midplane, which are all in the luminosity decay phase. The average time for radiation to reach quasi-steady state, i.e., a blackbody-like SED is about $6-8$min. This is shorter than the $t_{\rm ff}$ (Figure~\ref{fig:tffth_ddisk}) estimated using free-free opacity, potentially due to additional bound-free opacity \citep{nakar2010early,lovegrove2017very}. It is comparable to or longer than the time it takes for the star to pass the disk $H_{\rm disk}/v_{*}\sim0.4\rm min(R_{\rm ej}/R_{\odot})(v_{\rm *}/0.1c)^{-1}$ and ejecta expands $R_{\rm ej}/v_{\rm ej}\sim10\rm min (R_{\rm ej}/10^{12}\rm cm)(v_{\rm ej}/0.1c)^{-1}$, suggesting that SEDs might not have enough time to reach blackbody, consistent with SEDs shown in \textit{dynamic} runs.

The quasi-steady state SEDs share similar trend that the lower energy groups are near blackbody. The single temperature blackbody spectrum temperature and normalizations for backward and forward ejecta are $T_{\rm BB}=1.9\times10^{5}\rm K,~1.6\times10^{5}\rm K$, $L_{\rm BB,norm}=1.2\times10^{27},~4.0\times10^{27}$, with these temperatures roughly consistent with gas temperature near $\rthp$ calculated by frequency-integrated opacity.

The Compton tail approximately shows -2 slope, in the figure, we over-plot $\nu L_{\nu}\propto\nu^{-2.14}$ and $\nu L_{\nu}\propto\nu^{-2.36}$ for energy groups $h\nu>100.0$eV in forward and backward ejecta. The modification of SEDs by Compton scattering is more significant towards later times. The spectral energy enhancement for groups $h\nu>100.0$eV strongly correlates with the region of shocked disk instead of the ejecta, suggesting the bulk Compton may alter the shock structure and photon production. The bulk Compton scattering's effect will be explored in the future work with dynamical coupling between radiation and gas.

The relatively strong Compton effect we obtain from simulations are potentially affected by the assumption of fixing hydrodynamic evolution. The dynamic SEDs may not present such significant Compton tail as gas will be cooled by emitting photons, and the low energy component is likely to be less close to single temperature blackbody. Recently, \citet{vurm2025radiation} found Compton scattering can be an important source for soft X-ray emission in the ejecta from star-disk collision, especially in the higher incident velocity and higher gas temperature regime. Such thermal Compton effect near the shocked region may give rise to additional emission that we do not capture in the \textit{dynamic} runs.

\subsection{Dependence on Parameters}\label{subsec:multi_params}
We explore the effect of $H_{\rm disk}$, $\Sigma_{\rm disk}$ and $v_{*}$ with five \textit{dynamic} runs, which are listed in Table~\ref{tab:sim_multigroup}. For H1\_md\_v0.1c, H1\_md\_v0.07c, H3\_md\_v0.1c, the parameters correspond to their gray RHD counterparts (Table~\ref{tab:sim_params}). H3\_lld\_v0.1c and H3\_$\tau10$\_v0.1c do not have gray RHD counterparts. The former has $\Sigma_{\rm disk}=1.7\times10^{2}\rm g~cm^{-2}$ and $\tau_{\rm disk}=1.7\times10^{2}$, the latter has $\Sigma_{\rm disk}=52.3\rm g~cm^{-2}$ and $\tau_{\rm disk}\sim16$. The two runs are to test the interaction in the limit of low density, low optical depth disk.

\begin{table*}
\centering
\caption{Multigroup SED parameters}
\label{tab:sim_multigroup}
\begin{tabular}{ccccccccc}
\toprule

Name & $\Sigma_{\rm disk}(\rm g~cm^{-2})$ & $H_{\rm disk}/R_{*}$ & $L_{\rm peak,B}$ & $\sigma_{\rm rise,B}$ & $t_{\rm decay,B}$ & $L_{\rm peak,F}\tnote{2}$ & $-\xi_{\rm decay,F}$ & $\sigma_{\rm decay,F}$\\ 
\hline
H1\_ld\_v0.1c & $1.7\times10^{3}$ & 1.0 & $1.19\times10^{42}$ & 1.01 & 3.63 & $7.27\times10^{42}$ & -3.42 & 1.47 \\

H1\_md\_v0.1c & $1.7\times10^{4}$ & 1.0 & $1.22\times10^{42}$ & 0.85 & 4.00 & $2.79\times10^{43}$ & -2.05 & 0.61 \\

H1\_md\_v0.07c & $1.7\times10^{4}$ & 1.0 & $5.18\times10^{41}$ & 1.01 & 6.91 & $1.62\times10^{43}$ & -2.48 & 0.87 \\

H3\_md\_v0.1c & $1.7\times10^{4}$ & 3.0 & $1.73\times10^{42}$ & 1.17 & 6.54 & $3.95\times10^{42}$ & -4.11 & 2.82 \\

H3\_lld\_v0.1c & $1.7\times10^{2}$ & 3.0 & $5.91\times10^{41}$ & 1.19 & 4.39 & $1.18\times10^{42}$ & - & - \\

H3\_$\tau10$\_v0.1c & $5.2\times10^{1}$ & 3.0 & $2.81\times10^{41}$ & 1.48 & 4.00 & $1.77\times10^{42}$ & - & - \\
\bottomrule
\end{tabular}

\caption{Multigroup Light Curve Fittings Parameters: from left to right columns: Simulation name. $\Sigma_{\rm disk}(\rm g~cm^{-2})$: disk surface density. $H_{\rm disk}/R_{*}$: relative ratio of disk scale height and star radius. $L_{\rm peak,B}(\rm erg~s^{-1})$: the peak of $100.0\rm eV-1584.9\rm eV$ luminosity for backward ejecta. The rising light curve Gaussian fitting parameter $\sigma_{\rm rise}(\rm min)$ in Equation~\ref{eq:gaussianrise} for backward ejecta. The decay timescale for backward ejecta $t_{\rm decay,B}(\rm min)$. $L_{\rm peak,F}(\rm erg~s^{-1})$: the breakout luminosity in $100.0\rm eV-1584.9\rm eV$ for forward ejecta. The power-law decay fitting parameters $\xi_{\rm decay,F}$ and  $\sigma_{\rm decay,F}(\rm min)$ in Equation~\ref{eq:expdecay}. }
\end{table*}

\begin{figure}
    \centering
    \includegraphics[width=\linewidth]{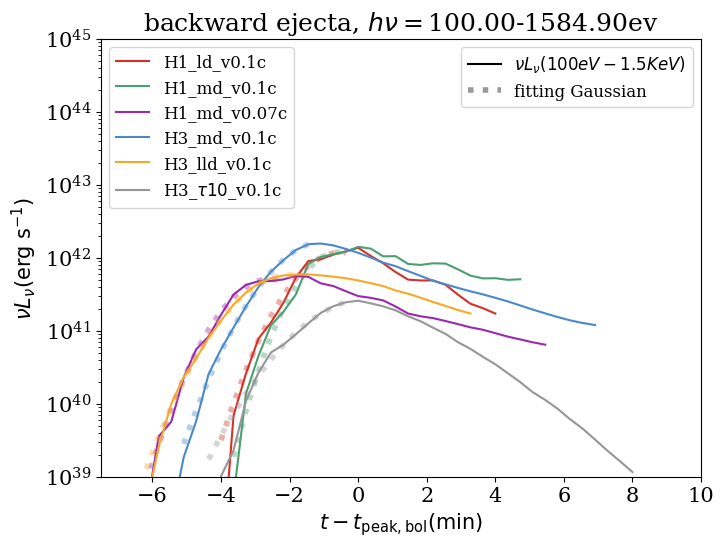}
    \includegraphics[width=\linewidth]{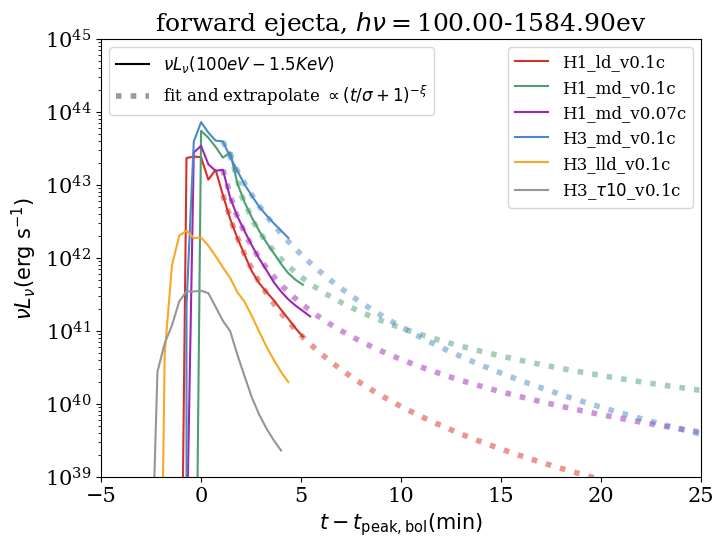}
    \caption{The upper panel: the backward ejecta luminosities summing over bands $h\nu=100.0$eV-$1584.9$eV. The lower panel: the forward dejecta luminosities summing over bands $h\nu=100.0$eV-$1584.9$eV. Green lines are the fiducial multi-group simulation H1\_md\_v0.1c ($H=R_{*},~\Sigma_{\rm disk}\sim10^{4}\rm g~cm^{-2}$); purple lines are fiducial disk but with lower incident velocity of $v_{*}=0.07c$; red lines are the simulation with lower disk surface density H1\_ld\_v0.1c ($\Sigma_{\rm disk}\sim10^{3}\rm g~cm^{-2}$); blue lines are disk with larger scale height H3\_md\_v0.1c ($H=3R_{*}$); yellow lines are a disk with larger scale height and low surface density ($H=3R_{*},~\Sigma_{\rm disk}\sim10^{2}\rm g~cm^{-2}$); gray lines are from the optically-thin disk with larger scale height ($H=3R_{*},~\tau_{\rm disk}\lesssim10$). The thick dotted lines in the backward ejecta rising and forward decay are the fitting according to Equation~\ref{eq:gaussianrise} and Equation~\ref{eq:expdecay}. The x axis time set the peak bolometric luminosity time as zero.}
    \label{fig:multi_fit_lc}
\end{figure}

We show the $h\nu=100.0$eV-$1584.9$eV bands light curve for these \textit{dynamic} runs in Figure~\ref{fig:multi_fit_lc}, which are primarily contributions from $100.0$eV and $251.2$eV bands. The first plot is for backward ejecta, the second plot is for forward ejecta. 

In the forward ejecta, we define the rise as the time corresponding to $\nu L_{\nu}=L_{\rm peak}/500.0$ to the peak luminosity $\nu L_{\nu}=L_{\rm peak}$, where typical band-dependent luminosity rises from $10^{40}\rm erg~s^{-1}$ to $10^{42}\rm erg~s^{-1}$. We fit a Gaussian function to the rise time in the backward ejecta $h\nu=100.0$eV-$1584.9$eV light curve: 
\begin{equation}\label{eq:gaussianrise}
    L(t)_{\rm rise,B}=L_{\rm peak,B}\exp(-(t-t_{\rm peak,B})^{2}/2\sigma_{\rm rise,B}^{2}),
\end{equation}
For the forward ejecta, we fit the decay by:
\begin{equation}\label{eq:expdecay}
    L(t)_{\rm decay,F}=L_{\rm peak,F}((t-t_{\rm peak,F})/\sigma_{\rm decay,F}+1)^{-\xi_{\rm decay,F}},
\end{equation}
where $L_{\rm peak,F}$ represents peak forward luminosity, $t_{\rm peak,F}$ is the time corresponding to peak luminosity. The power law decay of $L\sim (t/\sigma+1)^{-\xi}$ fitting format roughly estimates adiabatic cooling emission from expanding, optically thick ejecta \citep{vurm2025radiation}. We also define a decay time $t_{\rm decay,F}$ as it takes for the band-dependent luminosity to decrease by half of the peak luminosity

The fittings are shown as the dotted lines in Figure~\ref{fig:multi_fit_lc}. The peak $100.0$eV-$1584.9$eV luminosities $L_{\rm peak,B},~L_{\rm peak,F}$ and fitting parameters $\sigma_{\rm rise,B}$, $\sigma_{\rm decay,F}$, $\xi_{\rm decay,F}$ are listed in Table~\ref{tab:sim_multigroup}. We do not apply backward ejecta fitting to H3\_lld\_v0.1c and H3\_$\tau10$\_v0.1c, for which the decay light curve is not well approximated by adiabatic cooling model due to low $\Sigma_{\rm disk}$ and $\rho_{\rm ej}$.

We define peak time $t_{\rm peak,bol}$ as the time corresponds to maximum of bolometric luminosity, i.e. $\nu L_{\nu}$ summed over all photon frequency groups. For the backward ejecta, we compare the backward SED at the peak, 2.5min before the peak and 5.8min after the peak from different runs. When increasing $\Sigma_{\rm disk}$ in H1\_md\_v0.1c, or lowering velocity in H1\_md\_v0.07c, we find the backward SEDs are comparable to H1\_ld\_v0.1c (Section~\ref{subsec:multigroup_fiducial}). At 2.5min before the peak, the SED shape of H1\_md\_v0.07c is slightly harder with comparable total luminosity. The slightly harder SEDs shape is also reflected in Figure~\ref{fig:multi_fit_lc} backward light curve, its soft X-ray luminosity peaks earlier relative to bolometric peak compared to other runs. When increasing $H_{\rm disk}$ in H3\_md\_v0.1c, it shows higher spectral energy in all groups 2.5min before the peak due to slower rise, but the SEDs at the peak and 5.8min after the peak are similar to H1\_ld\_v0.1c. 

For the forward ejecta, by increasing $\Sigma_{\rm disk}$, H1\_md\_v0.1c shows slightly harder breakout SED than H1\_ld\_v0.1c, but the SED decay is similar to H1\_ld\_v0.1c. With higher $H_{\rm disk}$, H3\_md\_v0.1c has slower SED evolution than H1\_md\_v0.1c, the breakout energy is reduced but decays slower. Decreasing $v_{*}$ in H1\_md\_v0.07c, instead of significantly suppressing luminosity as implied by Equation~\ref{eq:ldiff} $L\propto v_{*}^{2}$, the forward SEDs are comparable with H1\_md\_v0.1c with slightly lower luminosity. To summarize, changing $\Sigma_{\rm disk}$ and $v_{*}$ has a moderate effect on SED shape within the tested parameter, increasing $H_{\rm disk}$ soften the SEDs by prolonging the rise and decay.

Several recent works summarized key observable features of current QPE candidates such as characteristic luminosity, duty cycle, flare energetics, and their correlations \citep[e.g.][]{nicholl2024quasi,mummery2025collisions,guo2025testing}. The observed typical peak soft X-ray luminosity for short duration events is roughly $L\sim10^{42}\rm erg~s^{-1}$, with a characteristic temperature of $kT\sim100$eV, giving a total released energy $E\sim10^{45-46}\rm erg$. In Table~\ref{tab:sim_multigroup} and Figure~\ref{fig:multi_fit_lc}, the backward ejecta peak luminosity of $kT\sim100$eV roughly agrees with the observed ones, but the duration is slightly shorter, resulting in lower total energy. The forward ejecta generally show higher luminosity of up to $L_{\rm peak,F}\sim10^{43}\rm erg~s^{-1}$, shorter rise times of less than a few minutes, and roughly consistent decay time, yielding released soft X-ray energy $E_{X}\sim10^{45}\rm erg$ and higher bolometric energy. Understanding physical processes that can soften and prolong the forward breakout emission is an interesting future topic, we also discuss briefly in Section~\ref{subsec:favored_condition}. 

The two lower optical-depth disks H3\_lld\_v0.1c and H3\_$\tau10$\_v0.1c show more significant differences. H3\_lld\_v0.1c produces ejecta with qualitatively similar morphology as H3\_md\_v0.1c, but substantially lower $\rho_{\rm ej}$. H3\_$\tau10$\_v0.1c do not show prominent ejecta, we discuss this run separately in Section~\ref{subsec:optically_thin_disk}. 

H3\_lld\_v0.1c has lower luminosity but longer rise and decay time than H3\_md\_v0.1c. In the backward ejecta, luminosity evolves slower in all photon frequency groups than other runs. Similar to H1\_md\_v0.1c, its pre-peak SED shape is harder and the UV to soft X-ray luminosity also peaks ahead of the bolometric luminosity peak. Comparing the forward SEDs at breakout, 2.2min and 4.5min after the breakout, H3\_lld\_v0.1c shows lower decay in bolometric luminosity. The SED shape is comparable to H3\_md\_v0.1c up to $h\nu=15.8$eV group, but the UV to soft X-ray luminosity is lower (Figure~\ref{fig:multi_fit_lc}), along with a weaker breakout emission.

\subsection{Collision with Low Optical Depth Disk}\label{subsec:optically_thin_disk}
In this section, we discuss results from H3\_$\tau10$\_v0.1c: the collision between a star and a low optical depth disk. We set the disk scale height $H_{\rm disk}=3R_{*}$, midplane density $\rho_{\rm disk, mid}=2.4\times10^{-10}\rm g~cm^{-3}$, corresponding to $\tau_{\rm disk}\approx16$ (with $\kappas$), which is significantly lower than the disks in previous sections with $\tau_{\rm disk}\sim 10^{3-4}$. The higher $H_{\rm disk}>R_{*}$ and lower $\rho_{\rm disk}$ may approximate sections in accretion disk that is magnetically elevated \citep[e.g.][]{begelman2017magnetically, mishra2020strongly} or flux-frozen \citep[e.g.][]{hopkins2023analytic}. If the QPE flares emerge from a relic TDE disk, however, it is less clear if the magnetic field is amplified strong enough to inflate the disk vertically at late time. This low density, puffy column of disk could also potentially represent the accretion flow during the TDE disk formation stage. Before a circularized, optically thick accretion disk forms, the scale height of accretion flow can be inflated due to outflow \citep[e.g.][]{ryu2023shocks,steinberg2024stream}. If the early time accretion flow extends to cross the EMRI star's orbit, they may experience similar star-disk collisions.

\begin{figure}
    \centering
    \includegraphics[width=\linewidth]{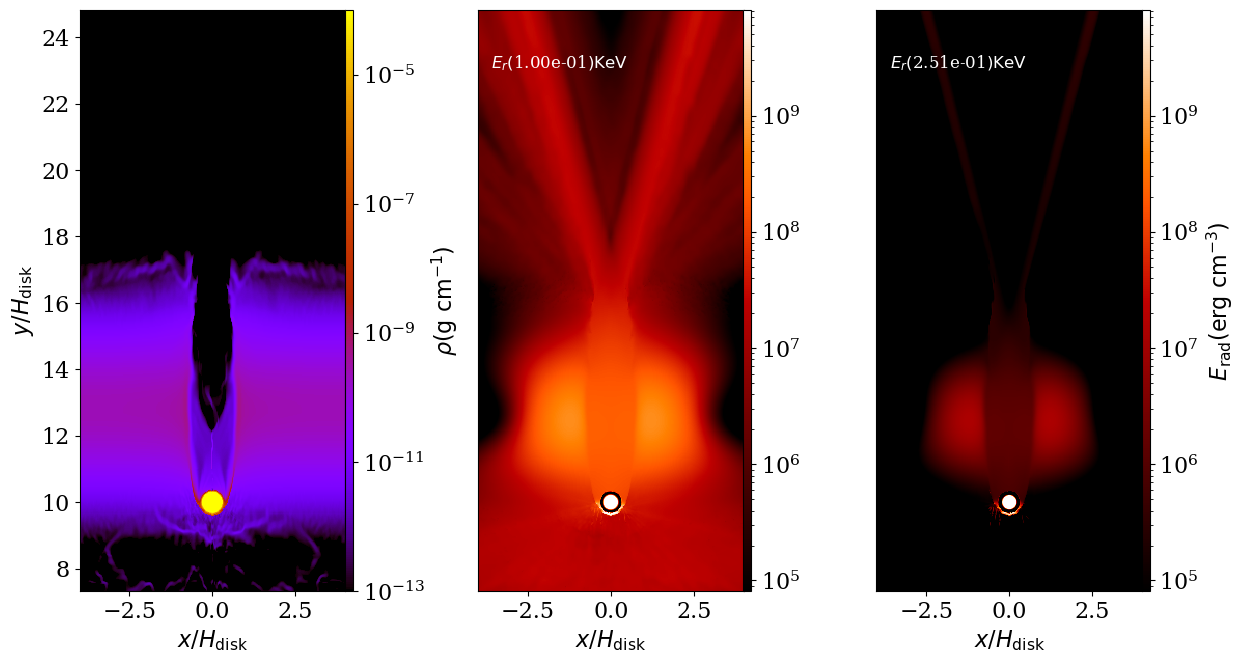}
    \caption{Density and radiation energy density snapshot from collision with a low optical-depth disk at t=3.5min. The left plot shows the gas density, the middle and right plot are radiation energy density map for $h\nu=100.0$eV and $h\nu=251.2$eV energy group. Similar to Figure~\ref{fig:hdisk_shock}, the ray effect of radiation energy density shows up in the backward ejecta direction, where the background gas is optically thin. }
    \label{fig:opticalthin_dens_er}
\end{figure}

In Figure~\ref{fig:opticalthin_dens_er}, we show a snapshot of gas density and radiation energy density in the $h\nu=100.0$eV and $h\nu=251.2$eV. A main difference compared to other simulations is the absence of backward and forward ejecta. When the star enters the disk, a similar bow shock forms in the disk and converts kinetic energy to radiation. Hot photons are generated near the shock front, enhancing local radiation energy density for $h\nu=100.0$eV and $h\nu=251.2$eV groups.

The star carves a low density channel in the disk. When it is at the midplane, gas temperature reaches $T\approx5\times10^{5}$K. Without optically thick ejecta to trap radiation, a small fraction of freshly produced photons quickly escapes through the optically thin channel, while the majority of photons diffuse through the disk with moderate optical depth. The gas cools to $T\approx10^{5}$K after the star passes the midplane. The small $\Sigma_{\rm disk}$ leads to lower radiation energy in the shocked gas. The radiation energy density roughly peaks in the $h\nu=39.81$eV band, and becomes insignificant for $h\nu\geq1.58$KeV bands (Figure~\ref{fig:opticalthin_dens_er}).

\begin{figure}
    \centering
    \includegraphics[width=\linewidth]{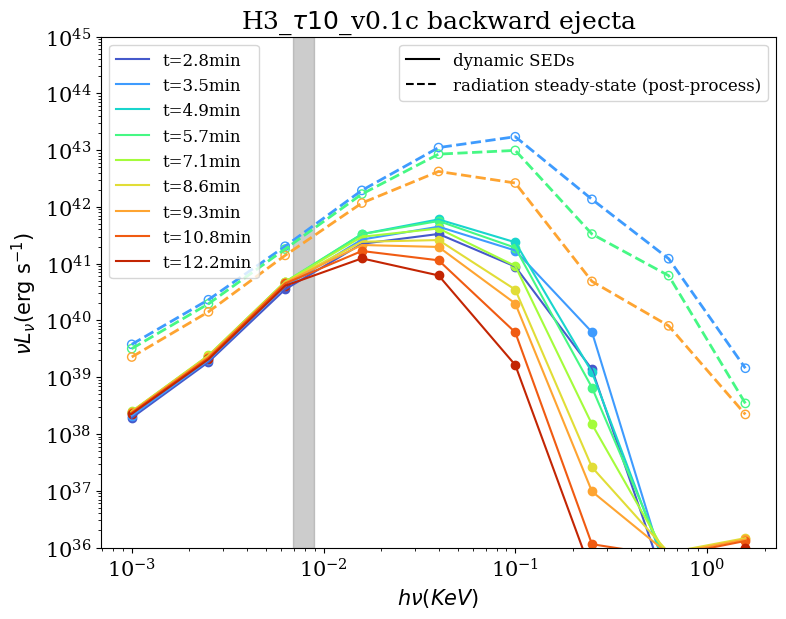}
    \includegraphics[width=\linewidth]{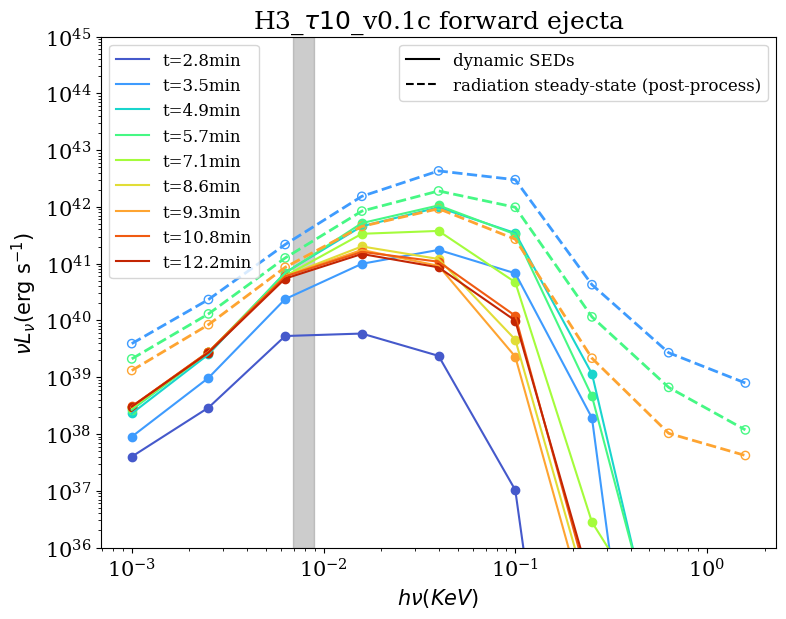}
    \caption{SED snapshots from the low optical depth disk.  The first plot is from the backward ejecta, the second plot is from the forward ejecta. The time labels are the time since the star is at the disk midplane. The solid lines and circles are calculated from \textit{dynamic} multigroup run. The dashed line and hollow circles are from \textit{post-process} multigroup run, i.e. the radiation quasi-steady state with Compton scattering. The gray shaded region labels the proposed UVEX far-UV band. }
    \label{fig:multi_spec_opticalthin}
\end{figure}

In Figure~\ref{fig:multi_spec_opticalthin}, the SED typically peaks in $h\nu=15.9$eV at earlier times and shift to $h\nu=39.8$eV later, generally lower energy than the optically thick disks. The spectral energy extends up to $h\nu = 251.2$ eV and cuts off in higher energy groups. The $h\nu=100.0-251.2$eV peak luminosity is $\nu L_{\nu}\approx 3\times10^{41}\rm erg~s^{-1}$, which is lower than typical QPE flares luminosity. The forward ejecta does not show a strong breakout emission as the bow shock travels through the low optical depth disk, so the forward and backward ejecta SEDs are less different. Due to the low optical depth and the short star-disk interaction time, the SED shapes deviate more from single temperature blackbody compared to optically thick disk with comparable $H_{\rm disk}$ and $v_{*}$. 

The quasi-steady state SEDs are affected significantly by bulk Compton scattering. Given the disk temperature $T\lesssim5\times10^{5}$K, thermal Compton has limited effect on the SEDs. For the backward ejecta, the Comptonized tail increases luminosity of $h\nu=100.0-251.2$eV by about two orders of magnitude in the decay phase, making the soft X-ray luminosity above $\nu L_{\nu}\gtrsim 10^{42}\rm erg~s^{-1}$, but still sharply truncates above $1$KeV. The quasi-steady state SEDs may overestimate the Compton scattering effect by allowing radiation evolves to quasi-steady state without feedback to gas thermal dynamics. 

The collision between a star and a low optical depth disk is usually not discussed in the context of QPE soft X-ray flares. From the simulations, we find such interaction can potentially produce UV emission with non-blackbody SEDs extends to soft X-ray band, although the soft X-ray luminosity is lower than typical QPE flares. But the collision with a nearly optically thin disk does not drive significant ejecta, photons are trapped by the disk materials. In Figure~\ref{fig:multi_spec_opticalthin}, we label the proposed UVEX \citep[][]{kulkarni2021science} far-UV band of $h\nu=6.5-8.9$eV as the gray shaded region. Depending on the recurrence time and whether bulk Compton can further increase the luminosity (the dashed lines), they might be relevant UVEX sources.

\cite{linial2024ultraviolet} proposed that when the photon production in the shocked disk is efficient and thermal equilibrium is achieved, the star-disk collision can produce thermalized flares with photon energy peaking in the UV band. The scenario can occur for a star orbiting relatively low mass SMBHs at wider orbital separations, intersecting a colder disk region with lower velocity  If the underlying disk is sufficiently cold, these UV flares can outshine the disk, giving rise to detectable quasi-periodic variability. Here, however, we demonstrate a different potential origin of UV flares when the disk surface density and optical depth are low, where thermalization is slow in the disk. The outcome SED is generally in lower energy, but is flatter in the optical-UV band than a blackbody SED, resulting in potential UV flare lasting tens of minutes (i.e., Fig.~\ref{fig:multi_spec_opticalthin}). Note that we do not consider the disk emission here, the flare may be detectable if it can outshine the low optical-depth disk's emission in relevant band.

In terms of extending the flare duration and increasing luminosity from the collision with a low optical depth disk, we speculate larger $H_{\rm disk}$ can increase star-disk interaction time and also prolong the photon diffusion time. Higher incident velocity can potentially enhance luminosity too. However, if the low-density, large-scale-height disk is more representative of the disk morphology farther from the black hole, then stars on orbits that cross these regions of the disk will have Keplerian velocities below $0.1c$. 
 
\subsection{Favored Condition for Short-duration QPEs}\label{subsec:favored_condition}

The multigroup runs show UV to soft X-ray luminosity and SED shape that are roughly consistent with the observed soft X-ray QPE flares \citep{chakraborty2024testing,arcodia2024more,nicholl2024quasi}. The main deviations are the rise time is shorter than some observed flares, and the SED evolution of backward and forward ejecta are different. Figure~\ref{fig:multi_fit_lc} fits forward ejecta rise time and extrapolates backward ejecta decay. 

The extrapolated soft X-ray luminosities of the forward ejecta suggest a decay timescales of approximately $30$ minutes, in promising agreement with some short-duration QPEs (e.g., GSN 069, RXJ1301.9+2747, eRO-QPE2, XMMSL1, eRO-QPE4: \citealt{miniutti2019nine,giustini2020x,arcodia2021x,arcodia2024more,chakraborty2021possible}). However, the rise time of the forward ejecta is substantially shorter than observed QPE flares. The backward ejecta rise time is longer than forward ejecta, but the rise and decay are still more asymmetric than typical observed flares. This finding is at odds with the observation that the odd/even QPE flares do not show clear morphological dichotomy.

Given the different light curve shape of the forward and backward ejecta, one possibility is that only one flare per orbit is observed, with the second flare largely obscured by the disk, such that the observed flares are consistently from either the forward or the backward ejecta. However, if only one flare per orbit is seen, the alternating pattern of long/short recurrence times would not be observed. If, alternatively, the disk is observed near edge-on, within an angle of order $\lesssim r_{\rm diff}/a_0$, both forward and backward ejecta clouds would be visible for both odd and even disk crossings, eliminating substantial morphological differences in their observed light curves.

Additional physical processes may act to prolong the breakout emission of the forward ejecta, so that the forward and backward will become more comparable and potentially enables two flares per star's orbit. In Figure~\ref{fig:multi_fit_lc}, we define the $t_{\rm rise,B}$ as the time it takes backward ejecta luminosity rises from $L_{\rm peak,B}/100$ to $L_{\rm peak,B}$. The forward ejecta has prompt breakout emission that typically lasts less than two minutes, so we do not list the rise time for forward ejecta. We list $\sigma_{\rm rise,B}$, $\sigma_{\rm decay,F}$ and $\xi_{\rm decay,F}$ in Table~\ref{tab:sim_multigroup} , and discuss potential directions to prolong the rise of $h\nu=100.00$eV-$1.58$KeV light curve, while maintain observable soft X-ray luminosity.

Increasing $\Sigma_{\rm disk}$ show moderate effect on short-term timing, but may extend the extrapolated decay (dotted lines in Figure~\ref{fig:multi_fit_lc}). In the $H_{\rm disk}=R_{*}$ runs, H1\_md\_v0.1c (second row) has similar $\sigma_{\rm rise,B}$ and $t_{\rm rise,B}$ as H1\_ld\_v0.1c (first row), their $h\nu=100.00$eV-$1.58$KeV luminosities are also similar. For the forward ejecta, increasing $\Sigma_{\rm disk}$ reduces $\xi_{\rm decay, F}$ and leads to flatter decay extrapolation. With $H_{\rm disk}=3R_{*}$, H3\_md\_v0.1c (fourth row) and H3\_lld\_v0.1c (fifth row) provide a more extreme contrast in $\Sigma_{\rm disk}$ with three orders of magnitude differences. In H3\_lld\_v0.1c, the low $\tau_{\rm disk}\sim100$ result in nearly optically-thin ejecta, the backward decay $t_{\rm decay,B}$ is shorter than H3\_md\_v0.1c, however, the backward rise $\sigma_{\rm rise,B}$ is still similar to H3\_md\_v0.1c.

Lowering $v_{*}$ slightly prolongs the rise and decay of backward ejecta, while the forward ejecta timing is insensitive to the change between $v_{*}=0.07c$ and $v_{*}=0.1c$. Comparing H1\_md\_v0.07c (third row) to H1\_md\_v0.1c (second row), we find $\sigma_{\rm rise,B}$ and  $t_{\rm rise,B}$ show moderate differences. The extrapolation in Figure~\ref{fig:multi_fit_lc} suggest the decay rate is also similar. Smaller $v_{*}$ leads to lower $v_{\rm ej}$ and lower $T_{\rm ej}$. The former increases $t_{\rm diff}$ in Equation~\ref{eq:tdiff}, the later increases opacity in the ejecta, the net effect will depend on the relative impact of these processes. For bolometric luminosity, smaller $v_{*}$ lead to longer $t_{\rm diff}$, but from multi-group results, the impact on $h\nu=100.00$eV-$1.58$KeV luminosity remains moderate. 

By containing the star and trapping the photons in the disk for longer time, larger $H_{\rm disk}$ tends to prolong and smooth the light curve. Increasing $H_{\rm disk}$ by three times from H1\_md\_v0.1c (second row) to H3\_md\_v0.1c (fourth row) extends $\sigma_{\rm rise,B}$, $t_{\rm decay,B}$ and $\sigma_{\rm decay,F}$. The $h\nu=100.00$eV-$1.58$KeV backward and forward luminosity are comparable between the two runs. However, the extrapolation suggests that H3\_md\_v0.1c may decay slightly faster than H1\_md\_v0.1c, with differences within order unity, up to 25min after the bolometric peak (Figure~\ref{fig:multi_fit_lc}). 

To prolong the luminosity evolution, increase $H_{\rm disk}$ or decrease $v_{*}$  represent a more favorable region in parameter space. Changing these parameters reflects variations in the disk structure that sets $H_{\rm disk}$ (and $\Sigma_{\rm disk}$), or variations in the stellar orbit that sets $v_{*}$. Meanwhile, the SEDs will response to the changes in $H_{\rm disk}$, $v_{*}$  or $\Sigma_{\rm disk}$, so these parameters can be not arbitrary tuned considering observable soft X-ray luminosity.

For example, decreasing $v_{*}$ will yield colder ejecta, lowering the bolometric and soft X-ray luminosity. Increasing $H_{\rm disk}$ , however,  leave moderate change the SED shape except for the softened breakout emission of forward ejecta. Increasing $\Sigma_{\rm disk}$ has more subtle impact on  SED shape. It leads to larger $\rho_{\rm ej}$ and hotter $T_{\rm ej}$. The former raises optical depth and suppresses UV to soft X-ray emission, while the later can enhance UV to soft X-ray emission by producing hotter photons and reducing absorption opacity. The resulting SED depends on the relative strength of the two effects. Another potential limit we discuss is the low optical depth disk, where the reprocess of hot, post-shock photon is not by the ejecta but from the disk itself. However, kinetic energy (Equation~\ref{eq:Eej}) will be reduced as disk density drops. A possible compensation maybe a larger $H_{\rm disk}$ to keep sufficient $\Sigma_{\rm disk}$, which can also increase the diffusion time in the disk. 

There are several important factors that can change emission and dynamics that we do not include in this work. For example, due to the disk precession, moderate eccentricity of stellar orbit and the disk rotation, the collision between the star and disk will be inclined instead of vertical \cite[e.g. see][]{chakraborty2024testing, arcodia2024more}. The inclination can change the relative speed between the star and the disk, the bow shock will also sweep larger volume of disk. \citet{tagawa2023flares} propose that this may prolong and soften the breakout emission and the breakout emission can largely contribute to the soft X-ray flares. 

In addition, the column of disk is simplified in our simulations. We do not consider the rotation, the turbulent velocity distribution, the magnetic field and realistic vertical thermal structure of disk. These factors may distort the ejecta and potentially leads to different optical depth along different lines of sights, which could change the emission property and SEDs.

Accounting for these potential effects that can extend rising timescale, ejecta cooling emission from parameters explored in this work is promising to explain the short-duration soft X-ray flares such as GSN069, eRO-QPE2 or eRO-QPE4 \citep[][]{miniutti2019nine,arcodia2021x,arcodia2024more}, whose flare duration is at the order of one hour. For longer duration events such as the repeating flares in AT2019qiz \citep[][]{nicholl2024quasi} or ANSKY \citep[][]{HernandezGarcia2025}, additional emission mechanism may need to be invoked. 

Another constraint for star-disk collision to produce QPE-like flare is the survivability of the star under the tidal influence of the black hole such as mass transfer or tidal heating. The duty cycle for QPEs shows relatively uniform distribution \cite[e.g.][and many more]{arcodia2025srg}, short-duration events correspond to shorter recurrence time and closer orbit to the black hole. For example, \citet{guo2025testing} suggest that for events with short recurrence time like eRO-QPE2, the star is likely to reside on the orbit that is comparable to the tidal radius of a solar-like star; thus, eRO-QPE2 may prefer a lower mass star. In this work, we modeled a solar-like polytrope, Equation~\ref{eq:ldiff} suggests that the luminosity $L\propto R_{*}^{2/3}$, a lower mass star may produce lower luminosity. The SED dependence may be less straightforward due to the non-linear dependence of opacity. 

Understanding the star's response to tidal force and repeating tidal heating from massive black hole and interaction with disk is optimal with careful treatment of self gravity and realistic stellar structure. Meanwhile, the tidal heating and ram-pressure stripping can also modify the star's evolution and lead to further mass transfer to the black hole \citep{law2020stellar,ryu2020tidal,lu2023quasi,linial2024tidal,yao2025mass,bandopadhyay2025repeated}, which may operates at longer timescale, we reserve the study for future work. 

We only model the interaction between a tightly bound star and an unperturbed disk. \citet{linial2023emri+} suggests as the star repeatedly impacts the disk, the star can be ablated due to the heat deposited to it. \citet{yao2025star} find that multiple interactions can unbind the stellar atmosphere and increase the effect cross section of the star, and the unbound gas can further fall back onto the disk and lead to shock-induced emission. Meanwhile, the disk gas is shocked and these perturbations may cause spiral acoustic waves in the disk. The collision also periodically perturbs the disk and injects heat to the disk, circularization shocks may emerge similar to what was proposed in \cite{lu2023quasi,Linial2024coupled}. 

Considering multiple interactions between star and disk, if a significant fraction of previous ejecta still remains near the disk and the disk precession is weak, it may change the local environment for the next collision by inserting a higher density region near the collision site. In Appendix~\ref{appendix:ejectaevolve}, we show that most volume of ejecta are likely to leave the collision region in less than one day. Some ejecta are still bound to the black hole and on mildly eccentric orbits that will return in a few days, which may induce secondary shock when they reach the disk. The majority volume of shocked disk gas will return to circular orbit in less than one day. 

We find that ejecta cooling emission alone prefers short duration event, which does not significantly dominated the growing QPE sample. In particular, the QPEs that associated with previous TDE activity \cite[e.g.][]{nicholl2024quasi, chakraborty2025discovery} seem to show longer duration. When searching QPEs that follows previous TDE-like activities, there can be observational bias that the longer duration events are more detectable by surveys given their longer recurrence times. The main challenge with applying the star-disk interaction presented in this work to such sources is their long duration and large total emitted energies. If the flare duration is primarily set by the diffusion time through the shocked ejecta, the long duration events impose stringent constraints on the disk surface density. 

A recent discussion of an additional emission mechanism that may play a role in long-period QPEs was presented in recent works \citep{yao2025star, linial2025qpes}. For example, \citet{linial2025qpes} studied the encounters between tidally elongated debris ablated from a star, and the underlying accretion disk. At periods of several days, the flare duration may be set by the arrival time of the stripped stellar debris, rather than the photon diffusion timescales of ejecta. While the two mechanisms - shocked disk material, as studied here, and shocked debris streams, as in \citet{linial2025qpes} – are closely linked, our current study focuses on the former. In sufficiently short period QPEs, stream-disk collisions may play a secondary role in terms of energetics, due to the higher optical depth of the ablated debris.

\section{Summary}\label{sec:summary}
We perform two sets of radiation hydrodynamic simulations to explore the collision between a star orbiting near a SMBH and a relic, low luminosity disk. The star is assumed to be solar-like, the disk scale height is comparable to stellar radius. The incident velocity is at the order of $v_{*}\sim 0.1c$, with the relative inclination between the star's orbit and the disk to be $90^\circ$. The set of gray RHD simulations (Section~\ref{sec:result}) focuses on the dynamics, the set of multi-group RHD simulations (Section~\ref{sec:result_multi}) investigate the photometric emission and the SED evolution. We focus on the collisions between an optically thick disk and a tightly bound star, but we also discuss the radiation signals from similar dynamical configuration for low optical depth disk (Section~\ref{subsec:optically_thin_disk}).

We find that as the star traverses the disk, a bow shock forms in the disk and local gas temperature increases to $T\gtrsim10^{6}$K. The collision is highly supersonic ($\mathcal{M}\gtrsim 10^2$), so that the interaction is governed primarily by the star's geometrical cross section, with no significant gravitational focusing. The bow shock drives two asymmetric ejecta clouds, from both sides of the disk (``forward'' and ``backward'' ejecta, Section~\ref{result:bow_shock}). As gas in the disk is compressed by the bow shock, the kinetic and radiation energy of disk material increases. When the radiation energy density peaks, it is roughly in equipartition with the kinetic energy (Section~\ref{subsec:restult_starvel}). After the ejecta emerges from the disk and expands, its internal and radiation energy are converted back to kinetic energy. The ejecta cooling emission is the primary source of flares in the simulations.

Both forward and backward ejecta clouds are optically-thick soon after emerging from the disk, and eventually become optically thin. The ejecta cooling emission produces peak bolometric luminosity $L_{\rm bol}\sim10^{42-43}\rm erg~s^{-1}$, with noticeable differences in light curve shape between the two ejecta clouds. The backward ejecta velocity is about half of the forward ejecta velocity, its mass is smaller by about one order of magnitude. Nonetheless, the backward ejecta luminosity is similar to the forward ejecta, to within a factor of two. The forward ejecta shows a unique, brief signal from shock breakout through the optically-thick disk. This brief phase is characterized by a luminosity of $L_{\rm bol}\gtrsim10^{42}\rm erg~s^{-1}$, lasting about $5-10$ min after the peak. The decay can be approximated by a power law, broadly consistent with the assumption of adiabatic expansion (Section~\ref{fig:lum_ddisk}), but the simulations do not capture the long-term decay.

We vary the collision parameters including the disk surface density $\Sigma_{\rm disk}$ (Section~\ref{subsec:restult_densdisk}), the incident velocity $v_{*}$ (Section~\ref{subsec:restult_starvel}) and disk scale height $H_{\rm disk}$ (Section~\ref{subsec:restult_hdisk}). We find that the typical ejecta temperature ranges from $10^{5}$ to $10^{6}$K. Higher $\Sigma_{\rm disk}$ produces ejecta with higher temperature, reducing absorption opacity (primarily bound-free) and leading to larger bolometric luminosity. An order of magnitude change in $\Sigma_{\rm disk}$ usually results in a factor of a few changes in peak luminosity.

The flare luminosity scales approximately as $v_{*}^{2}$. When varying $v_{*}$ we find qualitatively similar bolometric light curve shapes with different normalization. They also show a trend that higher $v_{*}$ yields a slightly faster evolving light curve. We explore the impact of the relative size of the star and the disk by varying the disk scale height. A smaller $H_{\rm disk}\sim R_{*}/3$ does not produce significantly different light curve and ejecta dynamics. Increasing scale height to $H_{\rm disk}\sim 3R_{*}$ leads to generally more symmetric backward and forward ejecta,  which is broadly consistent with the trend found in \citet{vurm2025radiation}: larger $H_{\rm disk}$ allows more disk gas to be entrained by the enhanced pressure, thus tends to drive more symmetric backward and forward ejecta. The luminosity evolves slower, the peak luminosity is comparable to those of $H_{\rm disk}\sim R_{*}$, the breakout emission is prolonged and softened by the larger $H_{\rm disk}$.

Parallel to the gray RHD simulation, we present another set of multi-group radiation hydrodynamics simulations (Section~\ref{subsec:multgroup_opacity}). The overall dynamics is relatively unaltered, but the total luminosity summed over all frequency groups can be higher than the gray RHD luminosity. The effect arises when the photon energy groups near soft X-ray show lower absorption opacity than the gray opacity for temperature and density relevant to the ejecta, which is an important factor to generate the soft X-ray excess in SEDs in addition to the thermal component.

We find that the typical ejecta cooling SEDs can be described as a blackbody component that peaks in 20-50eV, a non-thermal excess extends to 700eV and a potential Comptonized tail with approximately $\nu L_{\nu}\propto\nu^{-2}$ that extends beyond 1KeV, above which we do not model in the simulations (Section~\ref{subsec:multigroup_fiducial}). The forward ejecta breakout emission is hard in spectral energy and peaks $\gtrsim200$eV, but substantially brief and lasts less than five minutes. Without considering Compton scattering, the UV to soft X-ray luminosity arises where the absorption opacity in $\approx100-300$eV bands can be lower than the frequency-integrated opacity, leading to $\nu L_{\nu}\gtrsim10^{42}\rm erg~s^{-1}$ for band $h\nu>100.0$eV. The SED cuts off near 2KeV. 

The average time for radiation filed to reach quasi-steady state is longer than the star's crossing time through the disk, and it is comparable to the ejecta's expansion timescale, leading to SED deviation from a blackbody spectrum. A Comptonized tail may enhance the luminosity above 1KeV. Given the relatively low gas temperature $T\lesssim10^{6}$K, the Comptonized tail is likely associated with bulk fluid motion in the disk, where shock compression generate strong velocity divergence near the shock front and leads to SED hardening (Section~\ref{subsection:multi_compton}). 

The soft X-ray luminosity is comparable to some short-duration QPE flares. However, the simulation flares show more rapid rise timescale compared to the observed QPE flares, potentially due to the assumption of tightly bound star and idealized disk vertical structure. The simulation flare decay and adiabatic expansion extrapolation suggest that the $100.0$eV-$1.5$KeV luminosity will drop below $10^{40}\rm erg~s^{-1}$ after about 25 minutes from the peak (Section~\ref{subsec:favored_condition}), broadly consistent with decay timescale of short-duration soft X-ray QPE flares.

We discuss the effect of varying $\Sigma_{\rm disk}$, $v_{*}$ and $H_{\rm disk}$. Potential directions to extend the rise timescale while not significantly reduce soft X-ray luminosity include increasing $H_{\rm disk}$ and lowering the velocity $v_{*}$. Additional emission mechanisms may also act to prolong the rise timescale for each flare. The asymmetry of the forward and backward ejecta light curves (bolometric and band-dependent) suggest either we only see one flare per star orbit, or additional emission processes are required to reduce the differences between forward and backward ejecta, if we see two flares per star orbit. For example, an inclined collision may smooth the forward ejecta's breakout signal. An ablated star or a star with puffier atmosphere may change the ejecta morphology and evolution. 

We discuss the limit of a star colliding with a low optical depth disk. We find that the ejecta disappears when the disk vertical scatter optical depth is on the order of 10. The bow shock still produces hot photons in the disk and may give a dimmer, faster-evolving UV transient flare, but the spectral energy is too low to explain the soft X-ray QPE flares (Section~\ref{subsec:optically_thin_disk}). 

\begin{acknowledgments}
    We thank the frequent discussions about the QPE emission property with Daichi Tsuna, Tony Piro and Re'em Sari. We are grateful to the helpful suggestions we received in the early stages of this work from Philippe Yao, Sierra Dodd, Eliot Quataert, Wenbin Lu, Brian Metzger, and observational insights Joheen Chakraborty, Lisa Drummond, Riccardo Arcodia and Matt Nicholl. We thank the anonymous referee for useful suggestions that help to improve the work. XH is supported by the Sherman Fairchild Postdoctoral Fellowship at the California Institute of Technology. IL acknowledges support from a Rothschild Fellowship and The Gruber Foundation, as well as Simons Investigator grant 827103. This research benefited from interactions that were funded by the Gordon and Betty Moore Foundation through Grant GBMF5076. XH acknowledge use of the Lux supercomputer at UC Santa Cruz, funded by NSF MRI grant AST 1828315. Resources supporting this work were also provided by the NASA High-End Computing (HEC) Program through the NASA Advanced Supercomputing (NAS) Division at Ames Research Center. The Flatiron Institute is supported by the Simons Foundation. 
\end{acknowledgments}

\bibliography{ref}

\appendix 
\section{Polytropic Star}\label{appendix:polystar}
We adopt an approximated gravity potential from \citet{freytag2012simulations}:
\begin{equation}\label{eq:grav_phi}
    \Phi(r)=-GM_{*}(r_{0}^{4}+r^{4}/\sqrt{1+(r/r1)^{8}})^{-1/4}
\end{equation}
where $M_{*}$ is the mass of the star, $r_{0}$ and $r_{1}$ are the two smoothing parameters regarding the core and the outer envelope. 

\begin{figure}
    \centering
    \includegraphics[width=0.6\linewidth]{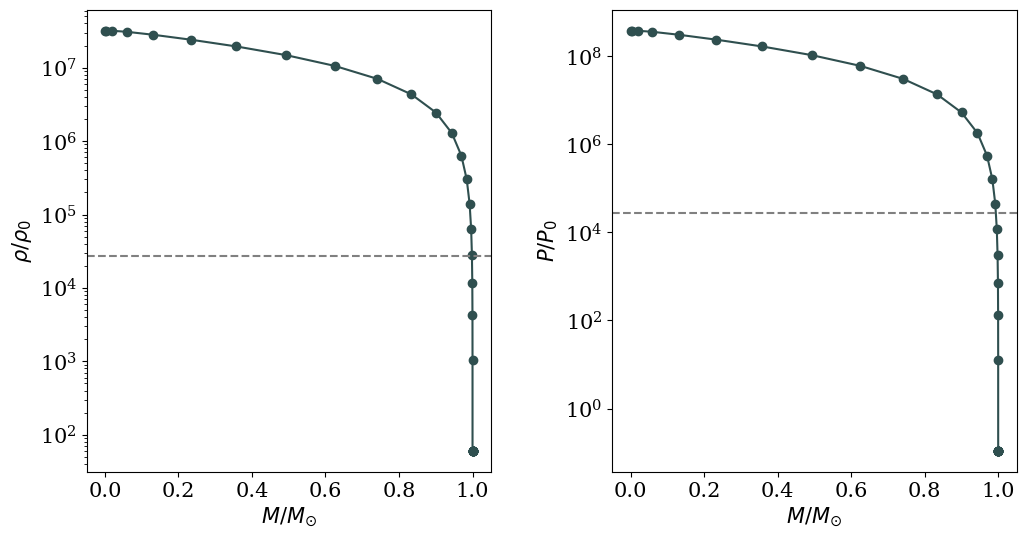}
    \caption{The numerical solutions for density (left) and pressure (right) of the polytrope star in the simulation. The density and pressure are normalized to density scaling unit $\rho_{0}=7.8\times10^{-7}\rm g~cm^{-3}$, pressure scaling unit $P_{0}=\rho_{0}v_{0}^{2}=8.1\times10^{7}\rm dyne~cm^{-2}$. The gray dashed lines show the ram pressure of the disk midplane. The dots show 32 interpolated points between $x/l_{0},~y/l_{0}=[0.001,0.313]$, roughly corresponding to our highest resolution region near the star, 11 points are closely overlapping at the coordinate $M/M_{\odot}=1.0$.}
    \label{fig:appendix_profile}
\end{figure}
We initialize the star as a polytrope with $n=3/2$. For gravity potential smoothing lengths, we adopt $r_{0}=\alpha\xi_{0}=0.790\alpha$, $r_{1}=\alpha\xi_{1}=3.505\alpha$, with canonical definition of $\alpha^{2}=\frac{(n+1)}{4\pi G}\frac{P_{c}}{\rho_{c}^{2}}$, where $P_{c}$ and $\rho_{c}$ are the central pressure and density. 

The pressure and density profiles are solved numerically from the Lane-Emden equations with the above gravity potential. Figure~\ref{fig:appendix_profile} shows the polytrope density and pressure profiles, with density scaling unit $\rho_{0}=7.8\times10^{-7}\rm g~cm^{-3}$ and pressure scaling unit $P_{0}=\rho_{0}v_{0}^{2}=8.1\times10^{7}\rm dyne~cm^{-2}$ as specified in Section~\ref{sec:method_set-up}. The gray dashed lines correspond to the disk ram pressure at the midplane $P_{\rm ram}=\rho_{\rm mid}v_{\rm disk}^{2}=\rho_{\rm mid}v_{*}^{2}$ for the fiducial disk. 

The profile is shallower than the Newtonian solution, equivalently the polytrope has a slightly more compact core and shallower envelope. As a result, the disk ram pressure can impact deeper in the polytrope pressure profile compared to the Newtonian solution. The spatial resolution near the star is about $0.01R_{*}$, which is larger than the scale height of the polytrope's outermost atmosphere. It is likely that the simulations overestimate the volume of stripped stellar atmosphere by disk ram pressure. But the total mass of ram pressure stripped atmosphere is moderate compare to the polytrope mass and thus less sensitive to the modified gravity potential.

When implementing the gravity source term $g(R)=-Gf(M_{\rm en}(R))$, we only apply the star gravity to the region of radius $R\leq R_{\odot}$. We justified in Section~\ref{result:bow_shock} that the Bondi radius of the star is sufficiently small compared to its radius, so the star-disk interaction is approximately geometrical instead of gravitational. To implement $g(R)$ in the Cartesian coordinate of our simulation, for each cell, we calculate its distance to the star center and linearly interpolate the density, pressure and $M_{\rm en}$. Effectively, $M_{\rm en}$ is fixed for the star during the simulation. Figure~\ref{fig:appendix_profile} suggest the disk ram pressure potentially impacts $<0.1\%$ mass in the disk, so the majority of star mass distribution is relatively unchanged during the interaction.

\section{Sampled One-dimensional Ejecta Profiles}
\label{appendix:1Dprofile}

Here we show more one-dimensional profiles sampled along the lines shown in Figure~\ref{fig:result_density_h1}. The backward lines (the brown lines) originate from $z_{\rm disk, mid}+2R_{*}$ with 10 uniformly separated angles $-45^{\circ}-45^{\circ}$ with respect to $x=0$. The forward lines (the blue lines) originate from $z_{\rm disk, mid}-2R_{*}$ with 10 uniformly separated angles $-60^{\circ}$ to $-10^{\circ}$, $10^{\circ}$ to $60^{\circ}$ with respect to $x=0$ to avoid the star. The thick solid line of each color shows the sampling lines that are discussed in Section~\ref{subsec:result_ejecta}. 

\begin{figure}
    \centering
    \includegraphics[width=0.5\linewidth]{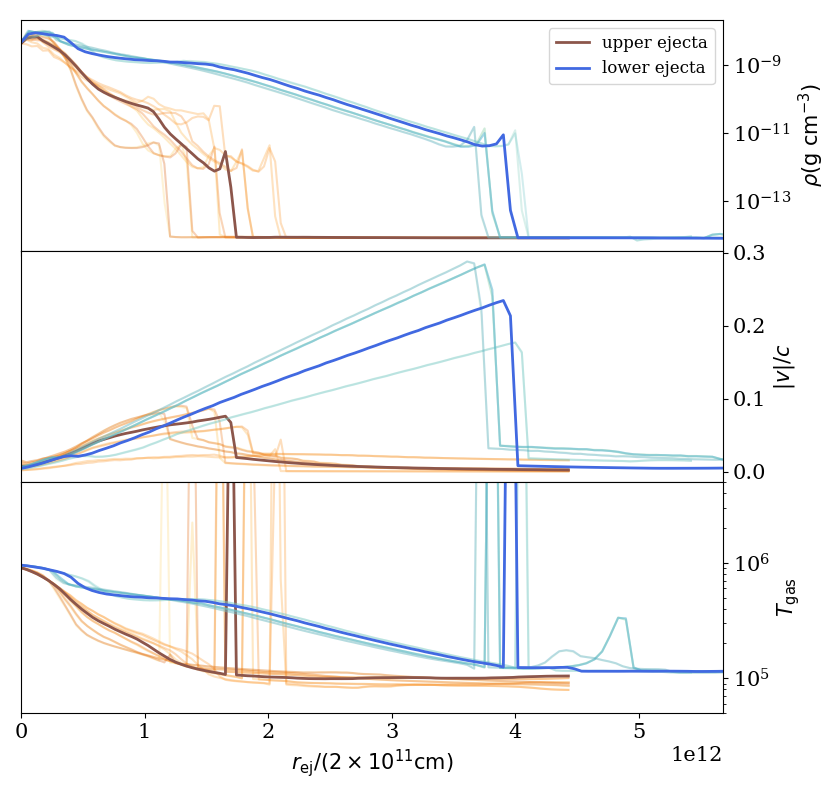}
    \caption{One-dimensional profiles of forward ejecta (light and dark blue lines) and backward ejecta (light and dark brown lines) in H1\_md\_v0.1c at $t=5.0$min since the star is at the disk midplane. The lines of sampled profiles are also shown in Figure~\ref{fig:result_density_h1}. The thick line with dark color is the representative line in each ejecta. From the first to the third row, we show density, velocity projected on the line and gas temperature profiles. The low density and high temperature feature in the brown lines are related to the small amount of low density gas at numerical floor post of backward ejecta. The high temperature feature in blue lines is the ``Zel'dovich spike''.}
    \label{fig:hdisk03_hydrolines}
\end{figure}

\section{Longer-Term Impact from the Ejecta}\label{appendix:ejectaevolve}
We randomly sample 1000 data points from the backward and the forward ejecta in H1\_md\_v0.1c at the time t=7.5min. The selection criteria for the sampling points include: its location is outside the disk (the vertical offset from disk midplane $|y-y_{\rm disk}|>5.6H_{\rm disk}$); it has more than $1\%$ concentration of disk passive scalar and has local gas density $10^{-4}<\rho_{\rm ej}/\rho_{0}<10.0$. Then we assume the collision happens at $100r_{g}$ from a $M_{\rm BH}=10^{6}M_{\odot}$ black hole. We use the sampled point's coordinate $x,~y,~z$ and velocity $v_{x},~v_{y},~v_{z}$ to calculate their location $x',~y',~z'$ and velocity $v_{x}',~v_{y}',~v_{z}'$ in the black hole centered frame, assuming a simple Newtonian transformation :
\begin{equation}
    x'=x+100r_{g},~y'=y,~z'=0.0,~v_{x}'=v_{x}, ~v_{y}'=v_{y},~v_{z}'=v_{\rm Kep}(R=100r_{g}),
\end{equation}
where $v_{\rm Kep}$ is the Keplerian velocity. We integrate the ballistic trajectories up to four days using the pseudo-Newtonian potential proposed by \citet{tejeda2013accurate}. The potential is designed to approximate apsidal precession near a spin-less black hole. It asymptotes to classic Newtonian potential at radius $\sim100r_{g}$. 

\begin{figure}
    \centering
    \includegraphics[width=0.95\linewidth]{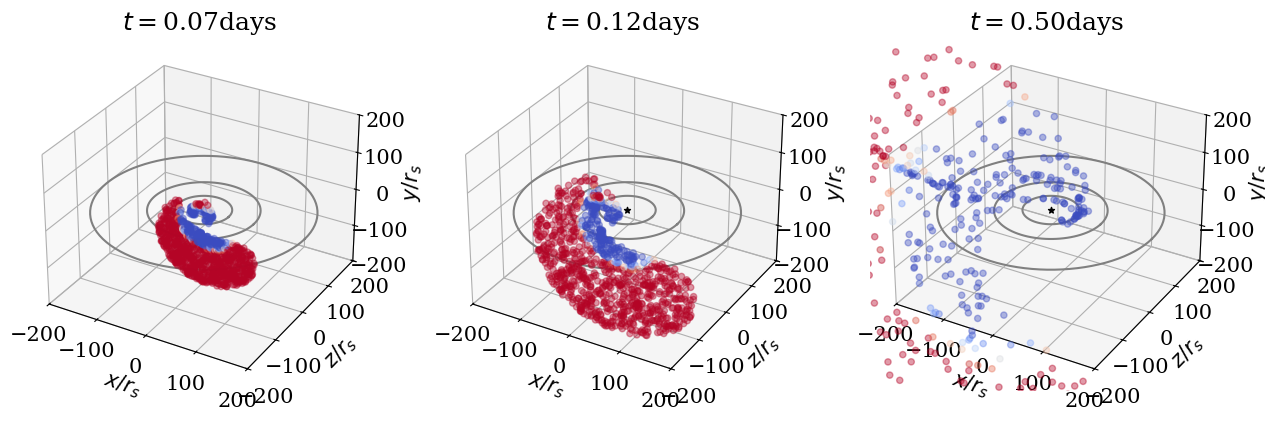}
    \caption{Locations of randomly sampled ejecta points that are extrapolated to 0.07 days (left), 0.12 days (middle) and 0.5 days (right) in the black hole centered frame. The red points are estimated to be with positive orbital energy (approximately unbound to the black hole), blue points are estimated to be with negative orbital energy (approximately bound to the black hole). The disk rotation is assumed to be clockwise (from positive z to negative z). The black hole location is at (0,0,0), marked by the black star. The three circles corresponds to $R=50r_{s}$ (where the star-disk collision occurs), $R=100r_{s}$ and $R=200_{s}$.}
    \label{fig:appendex_ej_scatter}
\end{figure}

\begin{figure}
    \centering
    \includegraphics[width=0.5\linewidth]{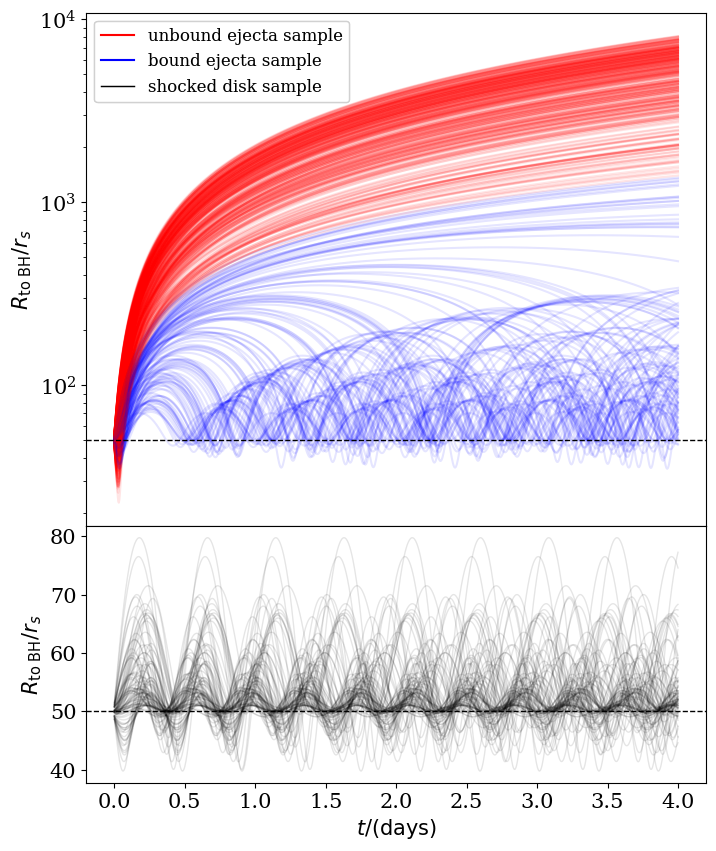}
    \caption{Evolution of sampled points distance to black hole extrapolated up to 4 days. The red line and blue lines represent points with positive orbital energy (approximately unbound) and negative orbital energy (approximately bound), randomly sampled from the ejecta. In the lower panel, the black lines are randomly sampled points from disk gas that is shocked and show enhanced radiation temperature. The black dashed line in both panels are the assumed initial collision position at $100r_{\rm g}$.}
    \label{fig:appendix_diskej_extrap}
\end{figure}

Figure~\ref{fig:appendex_ej_scatter} shows the extrapolated ballistic trajectories of the sampled data points. We label negative and positive specific total orbital energy (gravity potential and kinetic energy) by blue and red points. The disk is assumed to rotate clockwise. We find that including the disk Keplerian rotation can significantly change the extrapolated distribution of the sampled points compared to without the disk rotation. The positive energy points (red, approximately unbound) occupy larger volume but do not necessarily contain more mass than the negative points (blue, approximately bound), suggesting that the collision unbinds a fraction of the gas and making the remainder more gravitationally bound. The forward ejecta (negative $y$-coordinate) points has higher velocity so expands faster than the backward ejecta (positive $y$-coordinate). Majority of positive energy points escape from the collision site within a day, and may have moderate impact on the environment to the next collision.

The negative energy points (blue) are bound to the black hole and will return to the disk and on the timescale of their orbital periods. Figure~\ref{fig:appendix_diskej_extrap} tracks their distance to the black hole $R_{\rm toBH}/r_{s}$ with ballistic extrapolation. The assumed star-disk collision location $100r_{g}$ corresponds to the horizontal black dashed line in the plot. The red points are escaping from the black hole's gravity, so their $R_{\rm toBH}$ increases monotonically. The blue points are bound with pericenter close to original location $50r_{s}$, some of them gain larger eccentricity. An interesting future work may be exploring their orbital energy distribution as function of mass $dE/dM$ and estimating their mass fallback rate to the disk plane, similar to fallback rate calculation in TDEs. This requires three-dimensional simulation to model the disk rotation velocity during collision to better capture the ejecta energy redistribution. The readers are also referred to \citet{yao2025star} and \citet{linial2025qpes} for detailed discussion and calculation about fallback rate on to disk from the stripped stellar atmosphere and its implications to QPE flare emission source.

The lower panel shows the extrapolated $R_{\rm toBH}$ for another sets of points, for which we sampled as the shocked disk material. The selection criteria are inside the disk (the point has vertical offset from disk midplane $|z-z_{\rm disk}|<5.6H_{\rm disk}$), the point has a concentration of original disk material more than $1\%$ and the point has enhanced radiation temperature $T_{\rm rad}>10^{6}K$. The extrapolation suggests that the majority of shocked disk material gains moderate eccentricity and likely to rotate near its original orbit. When they damp the eccentricity in the disk flow, there may be secondary circularization shock and produce emission if the mass flux of eccentric shocked disk gas is significant.  

\end{CJK*}
\end{document}